\documentclass[10pt, conference, compsocconf]{IEEEtran}
\IEEEoverridecommandlockouts
\usepackage{cite}
\usepackage[hidelinks]{hyperref}
\usepackage{amsmath,amssymb,amsfonts}
\usepackage{algorithmic}
\usepackage{graphicx}
\usepackage{textcomp}
\usepackage{xcolor}
\def\BibTeX{{\rm B\kern-.05em{\sc i\kern-.025em b}\kern-.08em
    T\kern-.1667em\lower.7ex\hbox{E}\kern-.125emX}}

\usepackage{float}
\usepackage{subcaption}
\usepackage[utf8]{inputenc}
\usepackage{url}
\usepackage{graphicx}
\usepackage[linesnumbered,ruled,vlined]{algorithm2e}
\usepackage{tabularx}
\usepackage{balance}

\SetKwInOut{Input}{Input}
\SetKwInOut{Output}{Output\,}

\begin{document}
\title{An Intent-based Framework for Vehicular Edge Computing
\thanks{This work is partially supported by Australian Research Council (ARC) FT180100140 and DP230100081.}
}

\makeatletter
\newcommand{\linebreakand}{%
  \end{@IEEEauthorhalign}
  \hfill\mbox{}\par
  \mbox{}\hfill\begin{@IEEEauthorhalign}
}
\makeatother

\author{\IEEEauthorblockN{TianZhang He}
\IEEEauthorblockA{\textit{Faculty of Information Technology} \\
\textit{Monash University}\\
Melbourne, Australia \\
0000-0002-5472-7681}
\and
\IEEEauthorblockN{Adel N. Toosi}
\IEEEauthorblockA{\textit{Faculty of Information Technology} \\
\textit{Monash University}\\
Melbourne, Australia \\
adel.n.toosi@monash.edu}
\and
\IEEEauthorblockN{Negin Akbari}
\IEEEauthorblockA{\textit{Faculty of Information Technology} \\
\textit{Monash University}\\
Melbourne, Australia \\
negin.akbari@monash.edu}
\linebreakand
\IEEEauthorblockN{Muhammed Tawfiqul Islam}
\IEEEauthorblockA{\textit{School of Computing and Information Systems} \\
\textit{The University of Melbourne}\\
Parkville, Australia \\
tawfiqul.islam@unimelb.edu.au}
\and
\IEEEauthorblockN{Muhammad Aamir Cheema}
\IEEEauthorblockA{\textit{Faculty of Information Technology} \\
\textit{Monash University}\\
Melbourne, Australia \\
aamir.cheema@monash.edu}
}

\maketitle

\begin{abstract}
The rapid development of emerging vehicular edge computing~(VEC) brings new opportunities and challenges for dynamic resource management. The increasing number of edge data centers, roadside units (RSUs), and network devices, however, makes resource management a complex task in VEC. On the other hand, the exponential growth of service applications and end-users makes corresponding QoS hard to maintain. Intent-Based Networking (IBN), based on Software-Defined Networking, was introduced to provide the ability to automatically handle and manage the networking requirements of different applications. Motivated by the IBN concept, in this paper, we propose a novel approach to jointly orchestrate networking and computing resources based on user requirements. The proposed solution constantly monitors user requirements and dynamically re-configures the system to satisfy desired states of the application. We compared our proposed solution with the state-of-the-art networking embedding algorithms using real-world taxi GPS traces. Results show that our proposed method is significantly faster (up to 95\%) and can improve resource utilization (up to 76\%) and the acceptance ratio of computing and networking requests with various priorities (up to 71\%). We also present a small-scale prototype of the proposed intent management framework to validate our solution.
\end{abstract}


\section{Introduction}
The automotive industry is one of the fastest-growing industries. In recent years, the increased use of onboard microprocessors such as On-Board Units (OBUs) and sensors technology has led to technological advancements that enabled vehicles to provide various safety and driver assistance-related systems. For example, modern cars can autonomously drive to their destination, warn the driver of external hazards, and avoid collisions. However, many of these applications' growing demands for computational resources urge the use of the communication infrastructure and connection with Road Side Units (RSUs) to offload the heavy computational tasks. Moreover, vehicles are becoming increasingly connected, and V2X (Vehicle to Everything) communications enable vehicles to communicate with each other and the outside world, allowing applications to go beyond internal functions and provide improved awareness of impending events over a wider area. For many conventional connected vehicles, the network was only responsible for transferring data from vehicles to the cloud. However, applications are evolving into a highly distributed layer that resides directly within the network fabric. In fact, it is crucial to support many real-time applications and perform analytics at the edge as close as possible to the data source, e.g., real-time collision avoidance systems on autonomous vehicles. Consequently, Vehicular Edge Computing (VEC)~\cite{Liu2021, Raza2019} has become the mainstream paradigm to meet strict performance requirements such as response time and network bandwidth of real-time applications.

\begin{figure}[t]
	\centering
	\includegraphics[width=0.7\linewidth, trim={0cm 0.1cm 0cm 0cm}, clip]{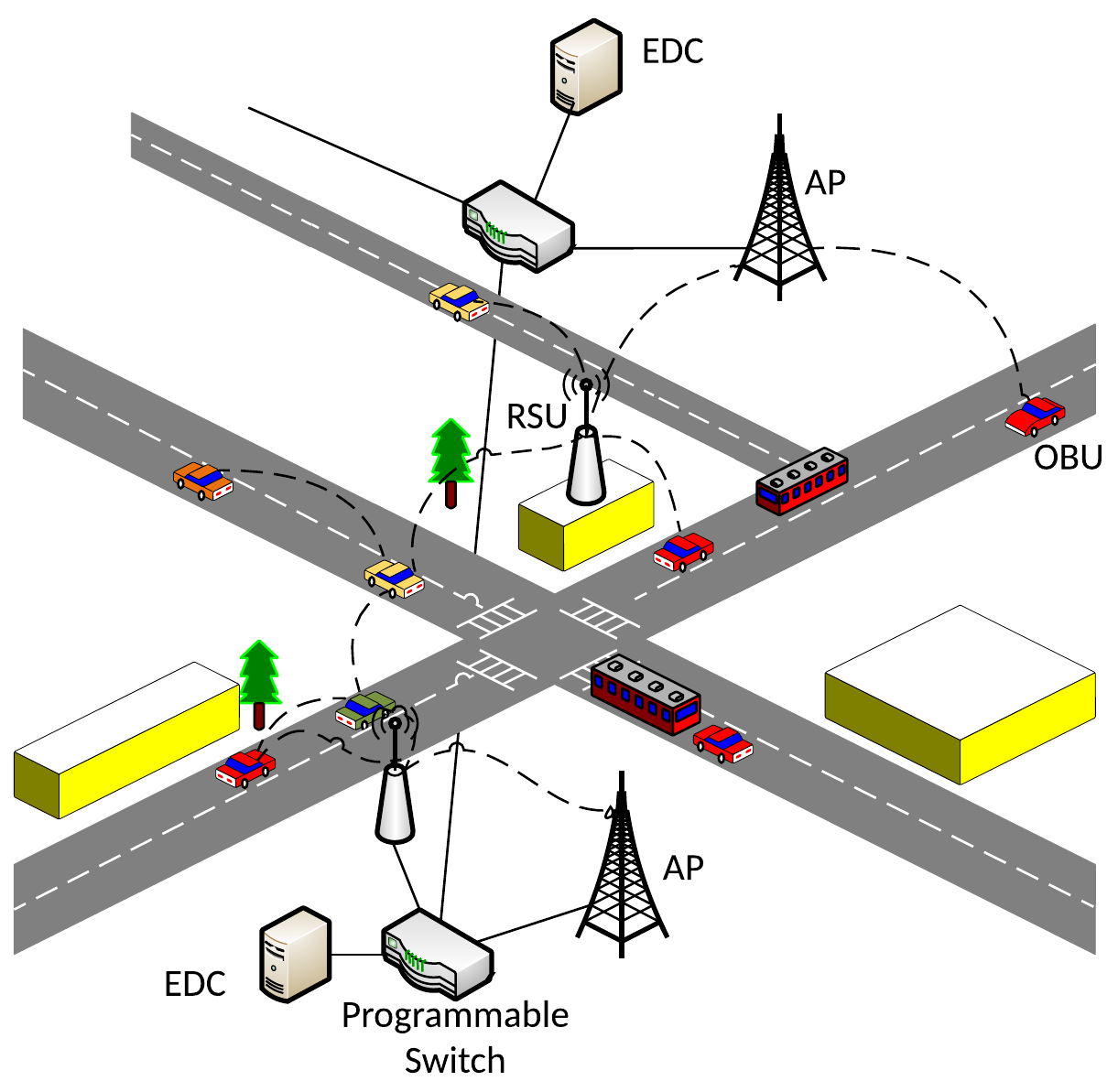}
	\caption{Vehicular Edge Computing Overview}
	\label{intent_vec_overview}
    \vspace{-0.5cm}
\end{figure}

With an increasing number of computational nodes, for example, cloud, edge devices, vehicles, RSUs, and compute-enabled network components, the networking operation domain in VEC is becoming more complex. Fig. \ref{intent_vec_overview} depicts one such VEC scenario where mobile vehicles may establish a network connection with other vehicles (inter-vehicle communication), RSUs, and access points (base stations). In addition, RSUs can be interconnected to other RSUs or to access points by either a wired or wireless network. Although each vehicle may have extra computing resources in its OBU, heavier computational workloads can be offloaded to the edge data centers (EDCs). Thus, it is challenging to efficiently orchestrate and manage the underlying networking and computing resources based on various service requirements in such a VEC environment. In other words, developing, deploying and operating applications in VEC environments is not trivial. Therefore, it is essential to create mechanisms to automatically capture the applications' deployment requirements (intents) to activate and assure them network-wide. Existing solutions try to address these issues mainly through service placement approaches and task offloading techniques without taking networks into consideration~\cite{7907225}. In this work, we aim to bridge the gap between the deployment requirements of VEC applications (business intent) and what the network delivers by building required algorithms considering both the computing and networking requirements of the applications. 

Software-Defined Networking (SDN) is a technology that helps tackle the network management and orchestration complexity~\cite{6994333} and has been widely used in VEC and Mobile Edge Computing (MEC) environments~\cite{8945202, 7140467}. SDN centralizes the network's control plane and provides automation, cost-efficiency, programmability, and greater efficiency for network management. 
In recent years, Intent-Based Networking (IBN) based on SDN is also emerging to automate networks further. It provides network intelligence by evoking a high-level intent, detecting potential deviations from that intent, and prescribing actions required to ensure that the intent is always satisfied, such as link rerouting and service reallocation via migration~\cite{he2022camig}. Based on IBN techniques, the application provides intents to indicate the desired network requirements, such as network bandwidth and end-to-end delay. 

Several industrial vendors, such as Cisco and VMWare, also focus on the IBN agility for edge network management. However, addressing only network resource requirements is insufficient in VEC. It should include both computing and networking requirements to guarantee end-to-end service delay. Furthermore, the network is dynamic at the edge, and failures or congestion can occur in switches, links, RSUs, EDCs, etc. A framework is required to automatically react to these issues and requirements to consistently and reliably maintain edge services operational. Following the industry trend, this work proposes an intent-based manager to orchestrate both networking and computing resources for VEC applications. Intent can be considered a high-level, abstract declaration for applications that describes their desired state or result~\cite{abhashkumar2017janus}. For example, a service provider may want an autonomous vehicle to maintain low-latency and high-throughput connections for its image processing service. An intent can be compiled into several service requests with resource requirements via existing network templates and policy languages, such as PGA, Kinectic, and Janus~\cite{abhashkumar2017janus} or NLP models~\cite{jacobs2018intent}.

To the best of our knowledge, this is one of the earliest efforts to install and support joint networking and computing intents for VEC applications. Current Virtual Network Embedding (VNE) algorithms~\cite{Yu2008, cheng2011vne, Fischer2013, gong2014vne, zhang2018vne, Thakkar2020} are inadequate for the intent installation problem in VEC environments for multiple reasons. Firstly, current VNE algorithms do not allow the allocation of multiple virtual nodes to the same physical node, negatively impacting resource utilization and the acceptance rate. Additionally, traditional virtual network requests (VNR) are treated as standalone, while an intent of a VEC application may need to be compiled into multiple VNRs. Current VNE algorithms also do not support adding location constraints as needed by VEC intents, making them incapable of handling the mobility aspect of the users/applications. Furthermore, current VNE algorithms do not consider computing requirements within the intent framework and models. Lastly, current VNE algorithms do not handle the installation of intents with priorities. Simply assigning priorities to the intents does not resolve the issue, as we also need to satisfy users' Quality of Service (QoS) requirements.

To address these problems, we propose an efficient online algorithm for intent installation with different priorities while considering both computing and networking-related properties. The key \textbf{contributions} of the paper are as follows:
\begin{itemize}
	\item We introduce the computing resource and location requests into Intent-Based Networking for VEC.
	\item We propose a priority- and location-aware algorithm for the installation and management of intents with different priorities.
	\item We compare our proposed algorithm with the state-of-the-art VNE algorithms in a large-scale simulation with real-world data sets and edge networks.
	\item We implement and evaluate the proposed intent-based edge computing with a real SDN controller on a Mininet emulation platform.
\end{itemize}

\begin{figure}[t]
	\centering
	\includegraphics[width=0.7\columnwidth, trim={0.6cm 0.6cm 0.6cm 0.6cm}, clip]{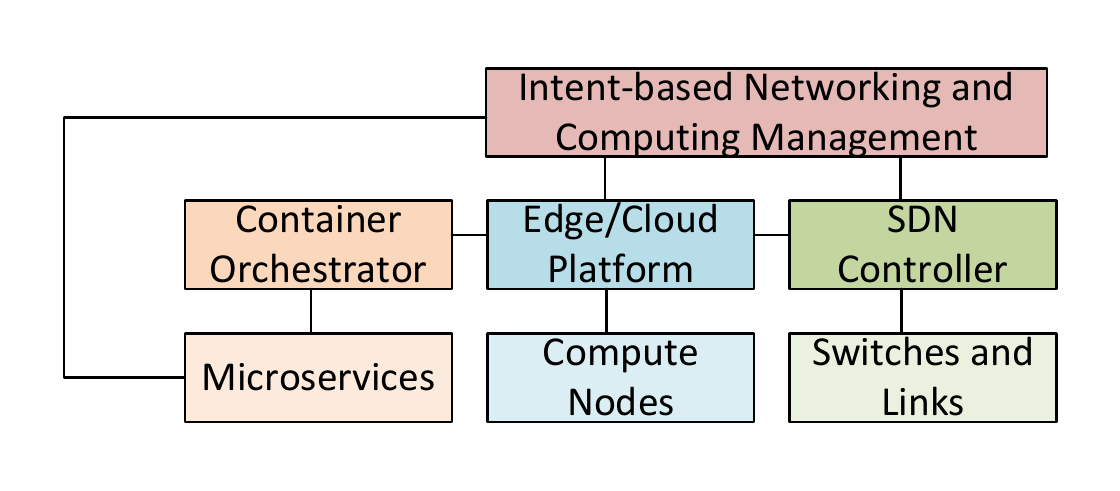}
	\caption{Intent management framework}
	\label{intent-mgt}
     \vspace{-0.5cm}
\end{figure}

\section{System Overview}
In this section, we first showcase the proposed intent management framework. Then we highlight the intent features available to the applications running in a VEC environment. Lastly, we discuss the intent life-cycle. 

\subsection{Intent Management Framework}

Current IBN frameworks only allow the mapping of application network resource  requirements. In this paper, our goal is to extend the intent framework to allow the expression of both compute and network elements along with application QoS. Thus, we need to holistically manage the edge/cloud platform for compute resources, orchestrate the container placements for micro-services, and  utilize the network controller to map the virtual network. As shown in Fig. \ref{intent-mgt}, by integrating SDN controller (e.g. ONOS or OpenDayLight), edge and cloud platform (e.g. OpenStack), and container orchestrator (e.g. Kubernetes), we propose an intent-based manager to cohesively orchestrate both networking and computing resources.

The applications' intents are submitted to the intent-based manager using a declarative manifest. The manager then extracts the requirements of each intent to check whether the intent can be installed with the existing compute and network resource capacity. For a successful intent compilation, the extracted compute requirements from the intent are transformed into a resource allocation task with the help of the edge/cloud platform and the container orchestrator. The network request is also transformed into a VNR and the manager instructs the SDN controller to install the network intent.

\subsection{Intent Features}

\begin{figure}[t]
	\centering
	\includegraphics[width=0.7\linewidth, trim={0.6cm 0.6cm 0.6cm 0.6cm}, clip]{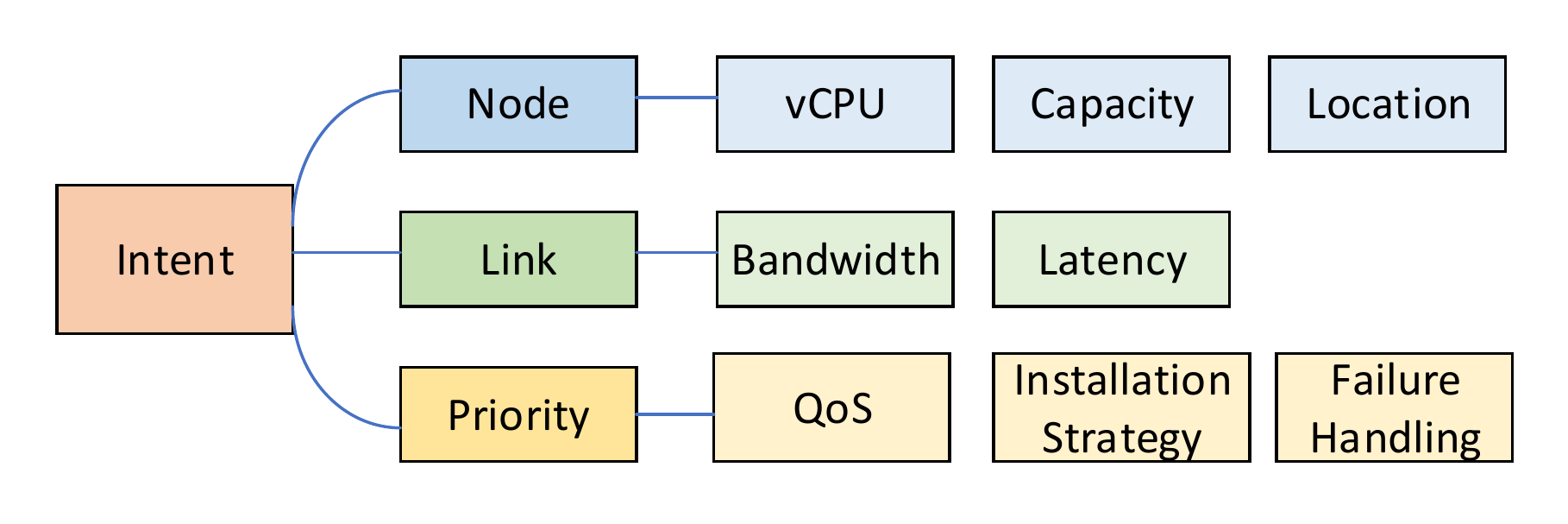}
	\caption{Intent features and constraints}
	\label{intent_properties}
    \vspace{-0.5cm}
\end{figure}

Fig.~\ref{intent_properties} shows the features available to the application to be expressed as per intent in the proposed intent management framework, which are categorized in three groups as follows:

\textbf{Node:} An application/service can choose a set of nodes (locations) where the service can be executed (location constraints). In addition, the compute resource requirements (e.g., vCPU, memory, or storage) for the service can also be specified (resource constraints).
    
\textbf{Link:} The application/service can choose the desired bandwidth requirement and express the minimum expected latency for the service to function correctly (network constraints).
    
\textbf{Priority:} As multiple services can be deployed across the system with competing and conflicting interests, intent from each service must have a priority specified while submitting the intent. Thus, depending on the intent priority and the QoS requirements, proper intent installation strategy and intent failure handling mechanisms can be followed.

\subsection{Intent Life-cycle}

The proposed intent framework allows applications to specify both their network and compute resource requirements. The intent manager accepts the intent specification requests and compiles them into installable intents that require some actions to meet the desired application state. Finally, when the actions are carried out in both the network and computing environments, some changes are made, for example, flow rules being pushed to switches or compute resources reserved to deploy a microservice on a node.

Fig. \ref{intent-lifecycle} depicts the complete life-cycle for the intents in the proposed framework. An intent can be in one of the following states: \textit{Ready, Active, Suspending, Failed, Withdrawn, Terminated} (represented in oval shapes in the figure). The intent manager takes the actions to make changes to the environment and satisfy intent requirements. These actions are represented with rectangular shapes. We provide a brief overview of the key actions in the life-cycle of the intents.

\begin{figure}[t]
	\centering
	\includegraphics[width=0.7\linewidth, trim={0.6cm 0.6cm 0.6cm 0.6cm}, clip]{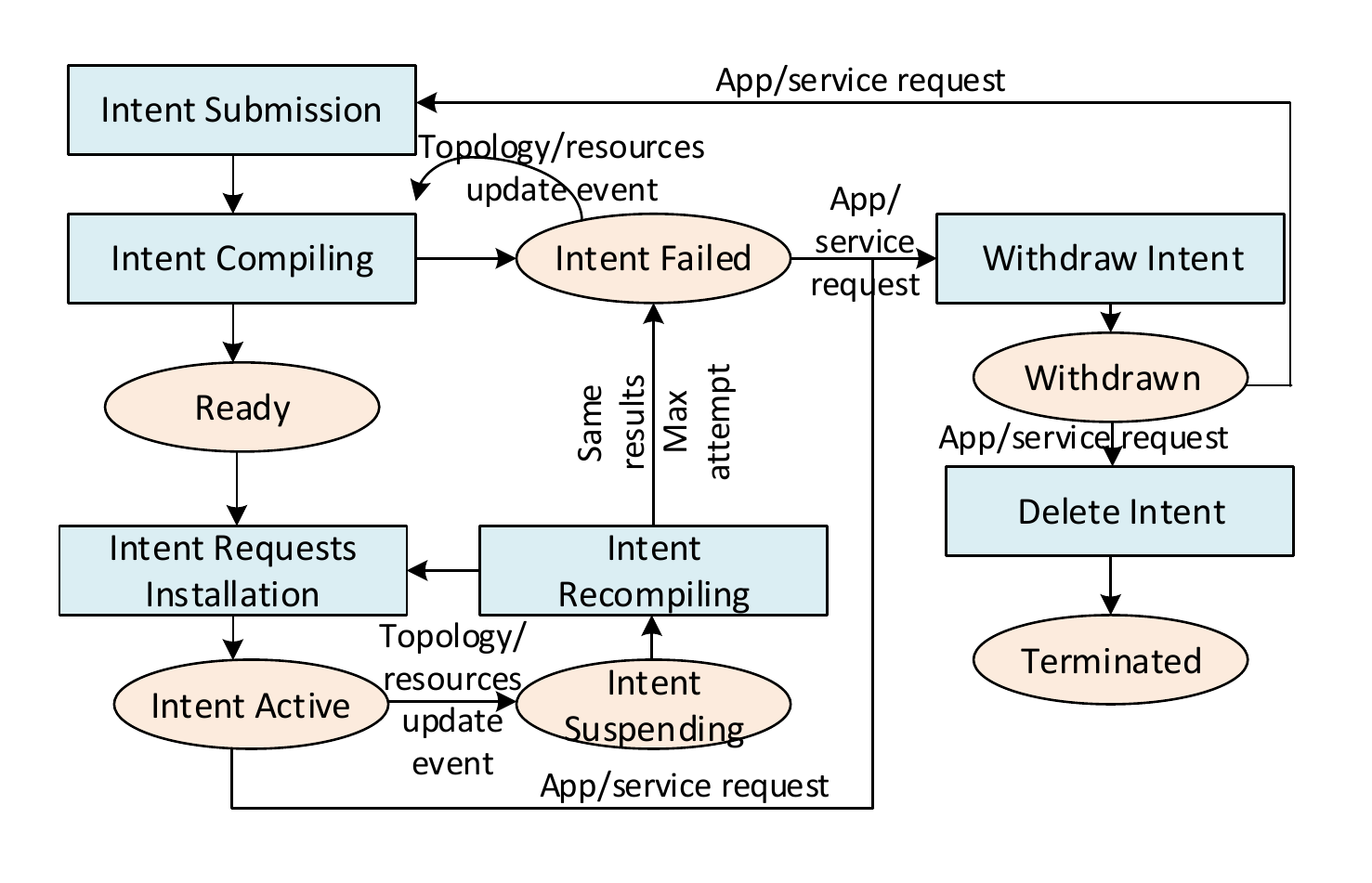}
	\caption{Intent lifecycle}
	\label{intent-lifecycle}
      \vspace{-0.5cm}
\end{figure}

\textbf{Intent Submission:} Intents are submitted by the applications/service providers to the intent management framework. Upon receiving an intent asynchronously, the intent manager transforms the intent into several compute and network requests which can be used for the intent compilation phase.

\textbf{Intent Compilation:} The resource discovery is made by the SDN controller and the VM/container orchestrator. The intent manager can communicate with them to get regular updates on network topology changes and the resource capacity of the compute nodes. Thus, if the network and compute resources are sufficient to handle the submitted intent, the intent will be compiled successfully, and the intent state becomes \textit{Ready}. Otherwise, the intent state goes to the \textit{Failed} state after exhausting the sufficient retry attempts.

\textbf{Intent Installation:} A \textit{Ready} intent can be installed in the system by reserving both the network and compute resources, such as compute and bandwidth for an application. When the intent is installed successfully, the state is changed to \textit{Active}.

\textbf{Intent Recompilation:} An \textit{Active} intent might be suspended by the application or due to a topology and resource update. In this case, the intent state is \textit{Suspending}, and it enters the Intent Recompilation phase. If the desired system state can be reached after a few recompilations attempts via link remapping or service migration~\cite{he2022camig}, then the intent re-enters an \textit{Installation} phase. Otherwise, the intent state is \textit{Failed}.

\textbf{Intent Withdrawal:} An application can request at any time to withdraw both an active or a failed intent. In this case, the system withdraws the intent, and if the intent is active, takes back all the allocated network and compute resources. A \textit{Withdrawn} intent can be resubmitted as a new intent based on the intent requirement template.  

\textbf{Intent Deletion:} While in the \textit{Withdrawn} state, the application can request to delete the intent requirement template entirely. At this point, the intent state will be \textit{Terminated}.

The retry threshold can be adaptively configured according to the current edge conditions. In highly dynamic environments, a small threshold can result in a high failure ratio. A large threshold, on the other hand, may cause a significant number of intents to be reinstalled, resulting in an increase in processing time. Furthermore, a large retry threshold can improve the acceptance rate for high-priority intents.

\section{System Model}

The phases of intent installations and contention resolutions between the intents can be translated into an\textit{ Online Virtual Network Mapping} or \textit{Virtual Network Embedding} (VNE) problem. The acceptance ratio and average resource utilization are critical parameters needed to be optimized.

\textbf{Vehicular Edge Computing:} VEC or a substrate network can be modeled as a weighted undirected graph $G(N,L,A_N,A_L)$ including the global network hardware, network links, edge devices and end users, where $N$ denotes the set of physical nodes and $L$ denotes the set of the physical network links. $A_N$ and $A_L$ denote attributes associated with nodes and links, respectively. For concise modeling, we consider CPU, memory capacities, and location constraints for node attributes, and bandwidth and latency for link attributes. We use $P$ to denote all loop-free paths within the VEC network and $P_{uv}^k$ to
denote $k$ shortest paths between node $u$ and $v$.

\begin{figure}[t]
	\centering
	\includegraphics[width=0.75\linewidth, trim={0.5cm 1cm 0.5cm 0.8cm}, clip]{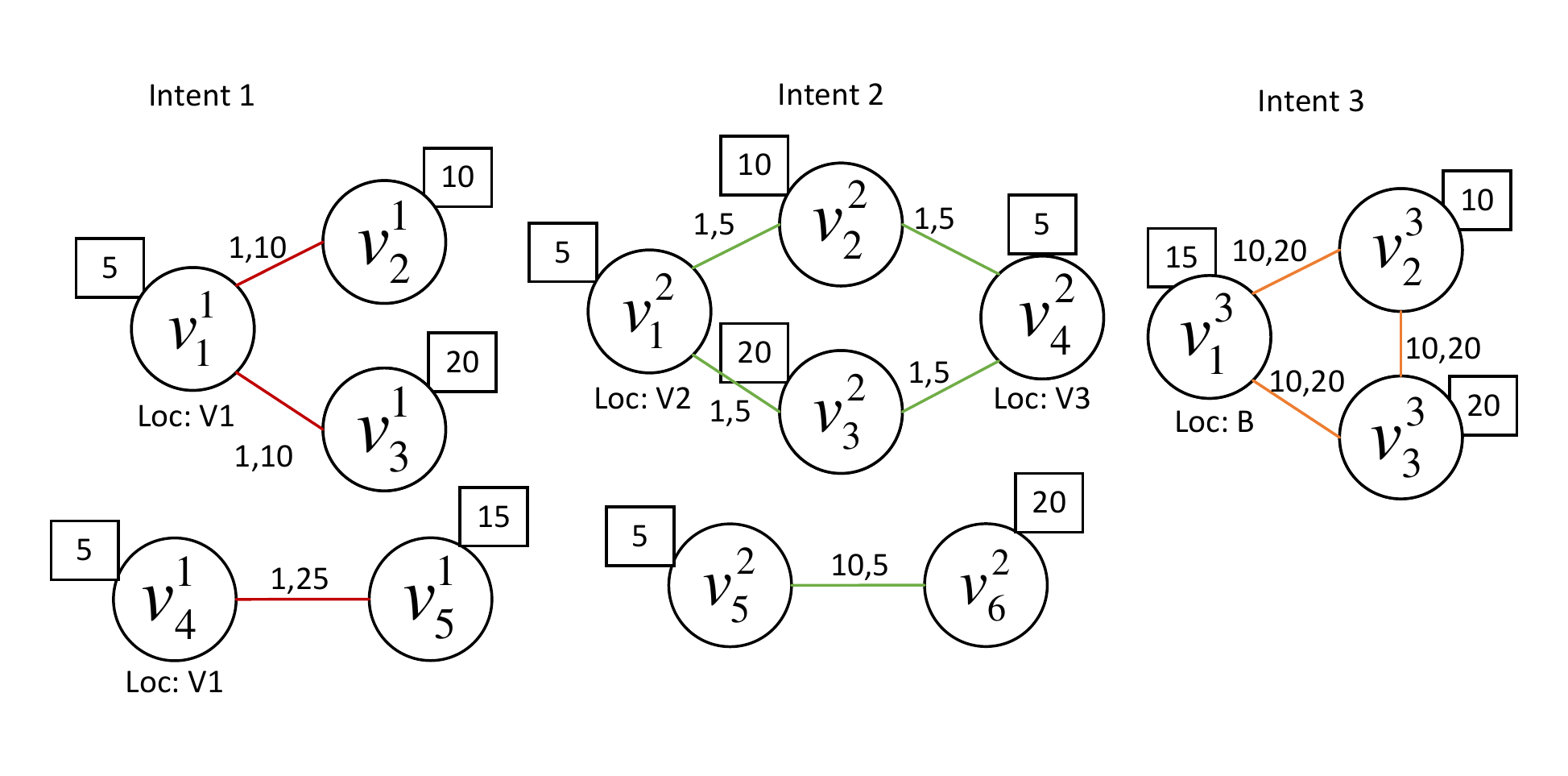}
	\caption{An example of intents and compiled requests}
	\label{fig: model1}
       \vspace{-0.3cm}
\end{figure}

\textbf{Intents and requests:}
Let $I^m$ indicate the intent submitted by the application/service $m$. Let $\pi^m$ denote the priority of Intent $I^m$. After the intent compiling process, $I^m$ can be compiled into several virtual networking and computing requests $R^m_i$ where $i$ denotes the $i$th requests of intent $I^m$.
As a microservice architecture, each compiled request of an intent can also be modeled as an undirected graph $R^m(N^m, L^m, C^m_N, C^m_L)$, where
$C^m_N$ and $C^m_L$ denote the set of node and link constraints, such as
the CPU, memory, and location constraints for virtual nodes, and bandwidth and latency for virtual links. For instance, in Fig.~\ref{fig: model1}, \textit{intent1} is compiled into two requests, where the computing requirement of virtual node $v_1^1$ is $5$. The bandwidth and delay requirement of the virtual link between $v_1^1$ and $v_2^1$ are 1 and 10, respectively. Virtual nodes $v_1^1$ and $v_4^1$ should be allocated in physical node $V1$, i.e., user node. As a result, if a mobile user reconnects to another EDC, virtual links associated with $v_1^1$ and $v_4^1$ need to be rerouted or other virtual nodes associated with these links might need to be relocated accordingly.

Location constraints $C^m_N(v) = n(v)$ of a virtual node $v$ can be divided into \textit{fixed} and \textit{mobility-related} location constraints. For the fixed location constraint, the virtual node location constraint is associated with a stationary physical node, such as a certain roadside unit, gateway, or edge data center. 
However, if location constraints are associated with mobile end-users such as autonomous vehicles or pedestrians, the location of the virtual node can change over time, i.e., $n(v)$ is a mobile node. The intent-based orchestrator actively monitors and maintains the intent installation for mobility-related location constraints due to the virtual node's location changes. 

\begin{figure}[t]
	\centering
	\includegraphics[width=0.7\linewidth, trim={0.6cm 0.6cm 0.6cm 0.6cm}, clip]{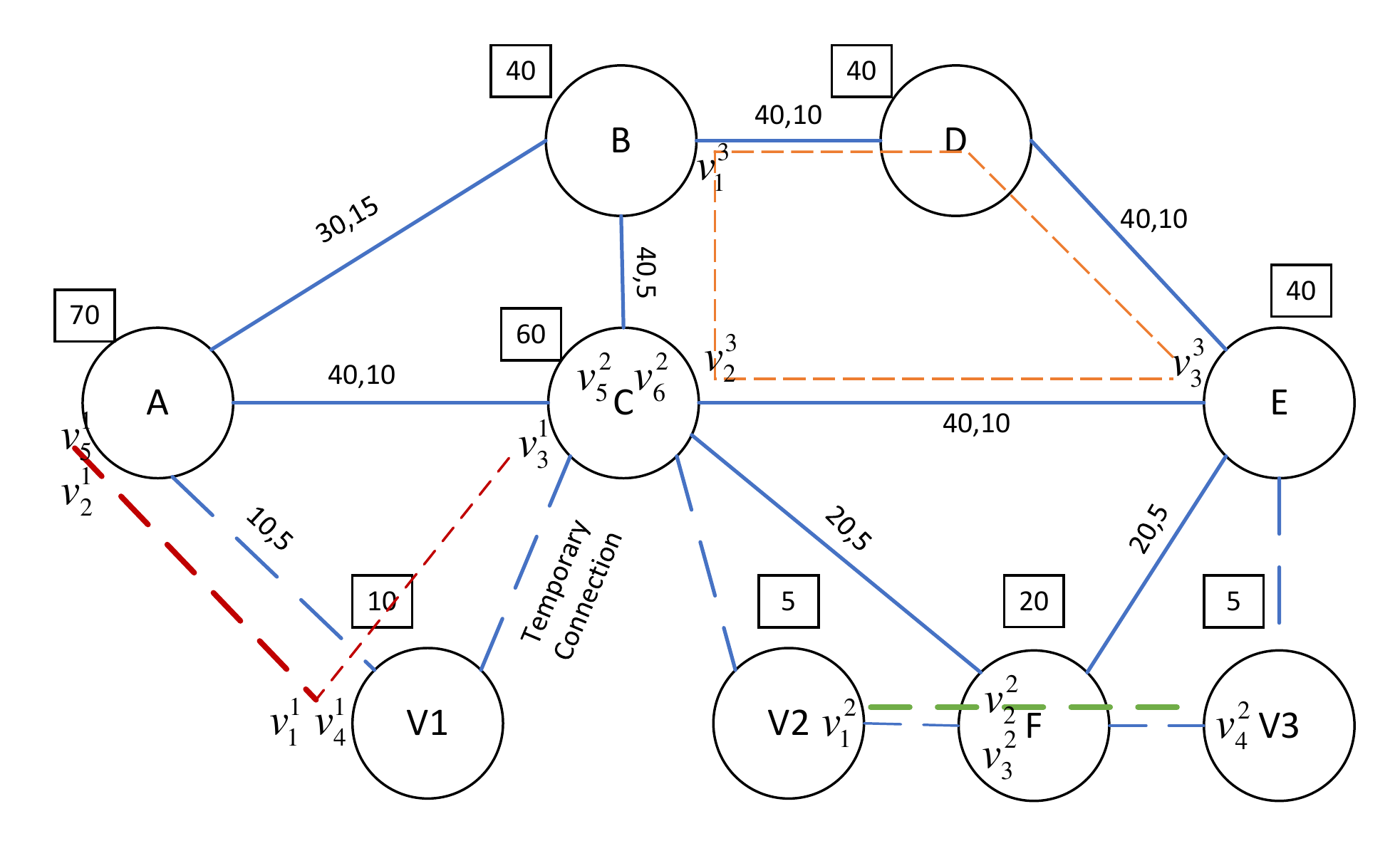}
	\caption{An example of intent installation on substrate network}
	\label{fig: model2}
         \vspace{-0.3cm}
\end{figure}

\textbf{Problem Description:}
The installation of a set of intents is defined by mappings:
$M{\{I^m\}}: \{R^m(N^m, L^m, C^m_N, C^m_L) \rightarrow G(N,P_{uv}^k,A_N,A_L)\}$,
from a set of $R^m$ to $G$, where $N^m \subset N$ is the node mapping and $L^m \subset P_{uv}^k$ is the virtual link mapping to the network path.
An intent installation is successful when its compute resource requirements and network resource requirements are all satisfied.
Fig.~\ref{fig: model2} illustrates a possible mapping for all requests of the three intents in Fig.~\ref{fig: model1}, installed on the substrate network. 

\textbf{Objectives:}
The objective of intent installation is to maximize the intent acceptance ratio owing to their priorities while efficiently utilizing both computing and networking resources.
Let binary variable $X^{m,t} = \{1,0\}$ denote whether intent $I^m$ is successfully installed at time $t$ or not and $X^{m,t}_i = \{1,0\}$ denote whether request $R^m_i$ of intent $I^m$ can be satisfied or not. When intent $I^m$ is not submitted, $X^{m,t} = 0$. Thus, the success of a general intent installation is defined as:
\begin{equation}
    X^{m,t} = X^{m,t}_0 \land X^{m,t}_1 ...
\end{equation}

Therefore, the long-term intent acceptance ratio is:
\begin{equation}
    \mathop {\lim }\limits_{T \to \infty } \frac{{\sum\nolimits_{t = 0}^T {\sum {{X^{m,t}}} } }}{{\sum\nolimits_{t = 0}^T {\left| {\left\{ {{I^{m,t}}} \right\}} \right|} }}
\end{equation}
where ${\left| {\left\{ {{I^{m,t}}} \right\}} \right|}$ is the total number of submitted intents at $t$.

We define three different priority levels for intents: \textit{high}, \textit{mid}, and \textit{low}.
With different priority levels, we define two installation semantics: 1) high and low priority intents are successfully installed only if all their compiled requests are satisfied and embedded $\forall X^m_i = 1$; and 2) a mid-priority intent allows its requests to be installed partially.

To quantify the acceptance ratio of mid-priority intents, we need to model the long-term request acceptance ratio, which can be formulated as:
\begin{equation}
    \mathop {\lim }\limits_{T \to \infty } \frac{{\sum\nolimits_{t = 0}^T {\sum {X_i^{m,t}} } }}{{\sum\nolimits_{t = 0}^T {\left| {\left\{ {X_i^{m,t}} \right\}} \right|} }}
\end{equation}
where ${\left| {\left\{ {X_i^{m,t}} \right\}} \right|}$ is the total number of requests at time $t$.

To the VEC provider, the cost of intent $I^m$ installation is modeled as the sum of total resource requirements:
\begin{equation} \label{eq: cost}
{\kappa ^m} = \alpha\sum\limits_{n \in {N^m}} {cp{u_n} + \beta \sum\limits_{n \in {N^m}} {me{m_n}} }  + \gamma \sum\limits_{l \in {L^m}} {{{b{w_l}} \mathord{\left/
 {\vphantom {{b{w_l}} {dela{y_l}}}} \right.
 \kern-\nulldelimiterspace} {dela{y_l}}}}
\end{equation}
where $\alpha$, $\beta$, and $\gamma$ are weights for resources in different categories valued by the VEC provider.
Thus, the revenue of intent $I^m$ installation for VEC provider at time $t$ can be formulated by:
\begin{equation}
    {\varepsilon ^{m,t}} = \sum {X_i^{m,t} \cdot \kappa _i^{m,t}}
\end{equation}

Similar to~\cite{cheng2011vne}, the revenue-to-cost ratio is used to quantify the long-term resource utilization:
\begin{equation}
    \mathop {\lim }\limits_{T \to \infty } \frac{{\sum\nolimits_{t = 0}^T {{\varepsilon ^{m,t}}} }}{{\sum\nolimits_{t = 0}^T {{\kappa ^{m,t}}} }}
\end{equation}

It is known that the general VNE problem is NP hard~\cite{Yu2008}. Thus, we rely on heuristics to practically solve the problem.

\section{Online Intent Management}
Existing solutions and algorithms~\cite{cheng2011vne, gong2014vne, zhang2018vne} of virtual embedding problem cannot be directly applied to the intent installation problem in VEC environments. In this section, we describe our proposed online intent management solution, which includes a priority-aware intent (PAI) installation algorithm and corresponding location-aware mapping (LAM) algorithm for intent-based VEC.
The intuitions behind our proposed priority-aware intent-based processing algorithm are:
\begin{itemize}
    \item Microservices (requests) with location constraints should be processed first to increase the acceptance ratio.
    \item Microservices with less complexity and resource demands that have less impact on other intent installations will be processed first.
    \item To increase the acceptance ratio of higher-priority intents and total resource utilization, intent requests with higher priority will be installed first. 
    \item If there is no request left for the highest installation level, we process compiled requests of all other intents.
    \item In the end, if there is no higher priority intent left for processing, we consider the requests with the lowest installation priority.
    \item There is a significant difference in processing and operation costs of intent reinstallation with virtual node relocation and merely remapping its virtual link~\cite{he2019mig}. The mapping may change due to the user mobility after a virtual node has been allocated in the mobile user node. Therefore, allocating other nodes that share virtual links in proper locations can reduce maintenance costs. 
\end{itemize}

\subsection{Priority Aware Intent Installation}\label{sec: PIA}
\begin{algorithm}[t]
\small
	\caption{Priority-Aware Intent Installation (PAI)}\label{alg: intent}
	\Input{ $G$ edge network graph with previous mapping information $M(t^{'})$}
	\Input{ time interval t, $\{I_{submit}\}, \{I_{suspend}\}, \{I_{fail}\}$}
	\ForEach{$ I \in \{I_{suspend}\}$}{
	\uIf{CheckLink(I) == Failed}{
	RemapPath(I);\\
	\uIf {status(I) == Failed} { $I_{fail} \gets I$}}}
	Compile($\{I_{submit}\} \cup \{I_{fail}\}$):\\
	Sort($\{I^{high}\}$); $\forall$ $I \in \{I^{high}\}$, InstallAll($I$);\\
	Sort($\{R^{mid}\}$); $\forall R \in \{R^{mid}\}$, InstallBest($R$);\\
	Sort($\{I^{low}\}$); $\forall I \in \{I^{low}\}$, InstallAll($I$);\\
\end{algorithm}

With different priorities, the intent contention resolution algorithm is the key component for IBN-based edge computing management and orchestration. At each time interval $t$, the priority-aware installation algorithm first checks intents associated with \textit{Suspending} event. If virtual link remapping cannot satisfy the intent, it changes the intent to \textit{Failed} state for reinstallation. We set the retry threshold for reinstallation from \textit{Failed} intent to 3.
To increase the acceptance ratio of intents with higher priorities and reserve sufficient resources for subsequent intents, we divide intents into two installation semantics and policies~(Algorithm \ref{alg: intent}):
\textit{\textbf{InstallAll($I$)}}: must satisfy all compiled requests of one intent.
\textit{\textbf{InstallBest($R$)}}: satisfy as many compiled requests as possible.
For each priority group, the intents $I^{high}$, $I^{low}$ and compiled requests $R^{mid}$ are sorted in ascending order based on the cost model Eq.~(\ref{eq: cost}).

\subsection{Location-Aware Mapping} \label{sec: LAM}

\begin{algorithm}[t]
\small
    \caption{Location-Aware Mapping (LAM)}\label{alg: lam}
    \Input{Requests $\{R\}$, Network $G$, Search depth $d$}
    \Output{$\{M(R)\}$ mappings of $\{R\}$}
    $\{R\} \gets \{R^{loc}\};$ or $\{R\} \gets \{R^{non}\};$\\
    sortRequests($\{R\}$);\\
    \ForEach{$R \in \{R\}$}{
    \ForEach{$v \in \{v^{loc}\}$}{
	$M(R) \gets$ mapNode($G, v, n(v)$);\\
	\If{$M(R)^v$ == null} {
	    status(R) = \textit{Failed}; goto line~\ref{alg: lam line: failed};\\
	}
    }
    
    sortNode($\{v^{non}\}$); 
    Q.enqueue(getMinNode($\{v^{non}\}$); \label{alg: lam line: sortvn}\\
    \While{$Q \neq \emptyset$}{
    $\{u\} \gets Q;$\\
        \ForEach{$u \in \{u\}$}{
            $M(R) \gets $ mapNode($G, u, M(R)$); \label{alg: lam line: mappednodes}\\
            \If{$M(R)^u == null$}{
            $M(R) \gets $ mapNode($G, u, sortNode(N_u^+)$); \label{alg: lam line: nearnodes}\\} 
            \If{$M(R)^u == null$}{status(R) == \textit{Failed}; goto line~\ref{alg: lam line: failed};\\}
            $\{u\} \gets $ Neighbors($R, u, M(R)$); Q.enqueue($\{u\}$); \label{alg: lam line: adjnodesnext}\\
        }
    }
    \If{status(R) == \textit{Failed}}{ \label{alg: lam line: failed}
    releaseMap(R);\\
    }
    }
    
\end{algorithm}

For intents and compiled requests embedding, we propose the location-aware microservice mapping (LAM)  algorithm by considering location constraints (Algorithm~\ref{alg: lam}). 
In addition to considering the computing and networking resource requirements of requests within intents, our proposed embedding algorithm also considers the location constraints, including fixed and mobility-related location constraints.
Traditionally, location constraints of virtual network requests are fixed, such as Virtual Network Function (VNF) location constraints of Service Function Chaining (SFC)~\cite{addis2015vnf} and host constraints due to data security or privacy.
For the intents with mobility-related location constraints, we allocate requests such that it minimizes the possibility of virtual node reallocation.

Following our installation semantics, the compiled requests $\{R\} = I^m$ of a high-level or low-level priority intent, and all compiled requests with mid-level priority $\{R\} = \sum I^m$ are processed sequentially.
LAM first allocates requests with fixed and mobility-related location constraints $R^{loc}$. 
Then, it allocates other requests without any location constraints $R^{non}$.
For each successful node mapping \textit{nodeMap}, the mapping result satisfies both node and link requirements. If a virtual link $l_{u,v}$ exists in $R$ and virtual node $v$ has been mapped to node $n(v)$, i.e., $M(R)^v = n(v)$, the virtual link is mapped based on $k$ shortest path between the testing node $n$ and $n(v)$. If any node mapping $M(R)^v$ is failed, the request mapping is also failed without further processing.

From steps 3 to 7, for each $R^{loc}$, virtual nodes  $\{v^{loc}\}$ with fixed and mobility-related location constraints $\{n(v)\}$ are mapped firstly. 
From steps 9 to 17, the remaining virtual nodes are mapped in a breadth-first search manner.
At step~\ref{alg: lam line: sortvn}, virtual nodes are sorted based on scores calculated in Eq.~(\ref{eq: virtual-node-score}); and the virtual node with minimum cost is selected ($u$) as the start node.
The virtual node score in a request $R$ is formulated as:
\begin{equation} \label{eq: virtual-node-score}
    S{(i')}^{R} = (\alpha' \cdot cpu + \beta' \cdot ram) \cdot k_{nn,i'}^{bw}
\end{equation}
where $cpu$ and $ram$ are normalized resource requirements of virtual node $i'$.
The ratio of $\alpha'$ and  $\beta'$ is based on the computing resource requirement of the selected virtual node $i'$,  ${\alpha'}/{\beta'}={{cpu}^{i'}}/{ram^{i'}}$.
$k_{nn,i}^{bw}$ is the average neighbor degree with bandwidth as the weight.
$ k_{nn,i}^{bw}= \frac{1}{s_i}\sum_{j \in {N(i)}}{bw}_{ij}k_j $, where $s_i$ is the weighted degree of node $i$, $N(i)$ is the set of node $i$'s neighbors, $k_j$ is the degree of node $j$ which belongs to $N(i)$. ${bw}_{ij}$ is the bandwidth of the edge (link) that connects node $i$ and $j$.

The node mapping procedure continues until the searching queue $Q$ is empty or a mapping is failed.
For each node mapping, already selected physical nodes in the request mapping $M(R)$ are tested first to minimize the link mapping cost at step~\ref{alg: lam line: mappednodes}. 
If there is no matching, candidate physical nodes within the search depth are sorted (Eq.~(\ref{eq: edge-node-score})) and tested for mapping at step~\ref{alg: lam line: nearnodes}.
At step~\ref{alg: lam line: adjnodesnext}, unexplored adjacent virtual nodes are enqueued based on the virtual node sorting for further mapping.
The candidates nodes $N_u^+$ for unmapped virtual node $u$ are the intersection set of edge nodes within search depth $d_{u,v}$ (the number of switch hops) of the current mapping node $n(v) \in M(R)$, 
\begin{equation}
  N_u^ +  = \bigcap\limits_{v \in M\left( R \right)} {N_{n(v)}^ + \left( {{d_{u,v}}} \right)}
\end{equation}
where $N_{n(v)}^ + \left( {{d_{u,v}}} \right)$ is all nodes within search range of node $n(v)$ and $n(v)$ is the mapping node of virtual node $v$ that $n(v): v \to n$.
The search depth is calculated as
${d_{u,v}} = d \cdot \left\lceil {{{Dela{y_l}} \mathord{\left/
 {\vphantom {{Dela{y_l}} {\mu}}} \right.
 \kern-\nulldelimiterspace} {\mu}}} \right\rceil $, $l_{u,v} \in L^m_i$ where virtual link $l_{u,v}$ exists, $\mu$ is delay coefficient ($\mu=10ms$) and $d$ is the search range coefficient ($d=2$).
If there is no mapped node, $N^+_u = G(N)$ for the first virtual node mapping.
When the request mapping is successful, the remaining physical resources are updated accordingly.

At step~\ref{alg: lam line: nearnodes}, the candidate physical nodes $N^+_u$ are sorted in a descending order based on the node score model (Eq.~\ref{eq: edge-node-score}).
The edge node $i$'s score in the substrate network is formulated as:
\begin{equation} \label{eq: edge-node-score}
    S(i)^{sub} = (\alpha \cdot cpu + \beta \cdot ram) \cdot k_{nn,i}^{bw}
\end{equation}
where the coefficients of remaining $cpu$ and $ram$ resources are $\alpha$ and $\beta$ ($\alpha + \beta = 1$). They are computed based on the remaining computing resources of all nodes within the search distance as:
\begin{equation}
    {\alpha}/{\beta} = {\sum_{j \in N^+_i(d)}{{cpu}^j}}/{\sum_{j \in N^+_i(d)}{{ram}^j}}
\end{equation}

Compared with traditional VNE solutions, we allow virtual nodes embedded in the same edge node to increase node utilization and reduce the network cost. In other words, each edge node can be a cluster of connected physical hosts. Multiple virtual nodes of the same request can be allocated in the same edge data center or the same physical host.
For the virtual link mapping, a remote edge node with longer paths results in reduced bandwidth resources for other intent installations. In other words, it may reduce the intent installation acceptance ratio when the network resources are limited.
Furthermore, most mapping algorithms consider the node and link mapping separately. However, virtual links of microservices at the edge are distance and latency-sensitive. Selecting all node locations without considering delays of a path can dramatically increase the reject ratio of intent installation. 
Therefore, we introduce search depth and distance parameters in node sorting to allocate requests in proximity.

\section{Performance Evaluation}

In this section, we compare the proposed intent installation algorithm with baseline algorithms based on the existing virtual network embedding algorithms, \textit{grcrank} (Global Resource Capacity)~\cite{gong2014vne}, \textit{rwrank} (Markov Random Walk PageRank)~\cite{cheng2011vne}, and \textit{nrmrank} (Node Ranking Metric)~\cite{zhang2018vne} in terms of intent acceptance ratio, resource utilization, and execution time. For a large-scale simulation, we utilize the real-world taxi GPS dataset in Shanghai (April 1, 2018)\footnote{\url{ http://soda.shdataic.org.cn/download/31}} and the locations of base stations from Shanghai Telecom.\footnote{ \url{http://sguangwang.com/TelecomDataset.html}}

\subsection{Experiment Configurations}

\begin{figure*}[t]
\centering
\begin{subfigure}[t]{.24\linewidth}
	\centering
	\includegraphics[width=\linewidth, trim={0.2cm 0.3cm 0.2cm 0.2cm}, clip]{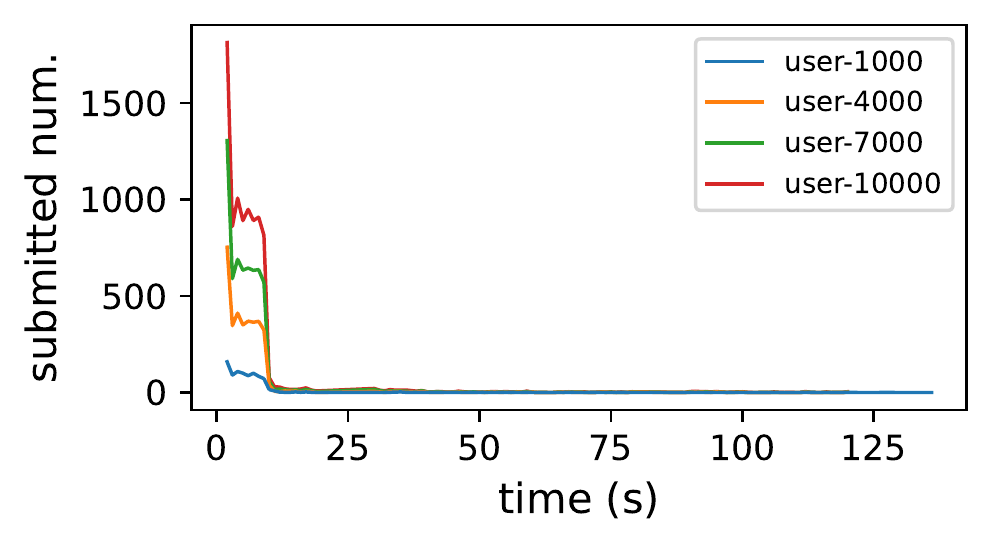}
	\caption{submitted num. of intents along time}
	\label{fig: input-submit}
\end{subfigure}
\hspace{0.1em}
\begin{subfigure}[t]{.24\linewidth}
	\centering
	\includegraphics[width=\linewidth, trim={0.2cm 0.3cm 0.2cm 0.2cm}, clip]{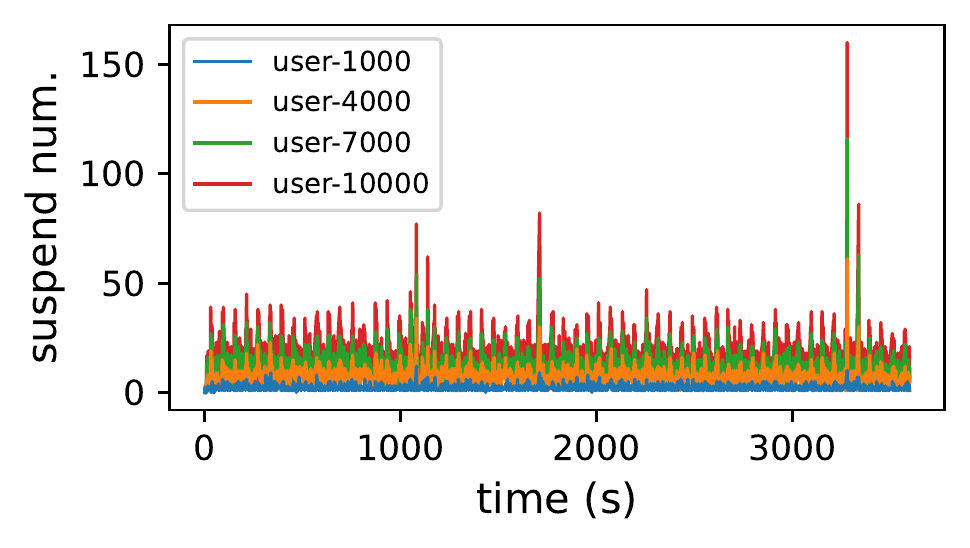}
	\caption{suspended num. of intents along time}
	\label{fig: input-suspend}
\end{subfigure}
\hspace{0.1em}
\begin{subfigure}[t]{.25\linewidth}
	\centering
	\includegraphics[width=\linewidth, trim={0.2cm 0.2cm 0.2cm 0.2cm}, clip]{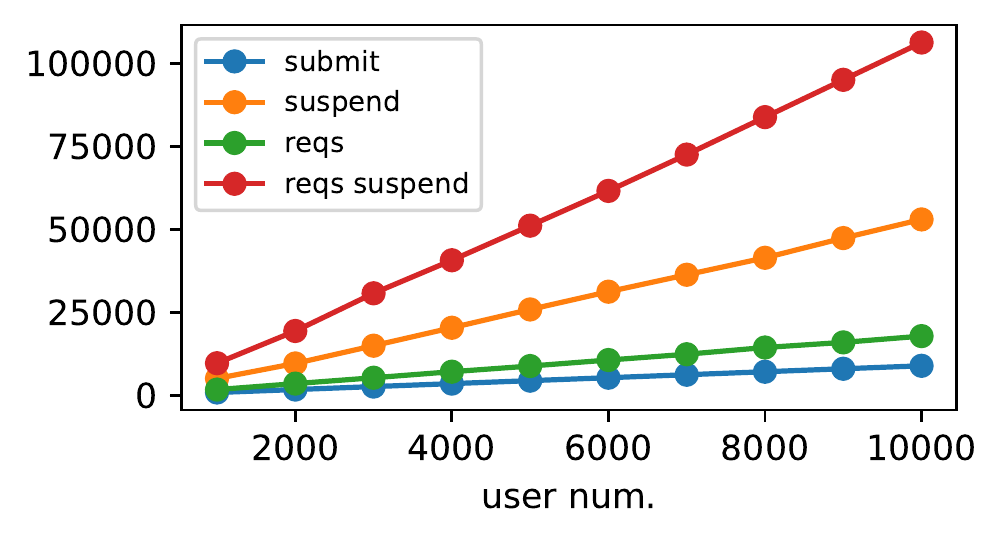}
	\caption{total num. of submitted and suspended intents and requests}
	\label{fig: input-num}
\end{subfigure}
\hspace{0.1em}
\caption{Intent data statistics with various amount of users}\label{fig: input}
\vspace{-0.3cm}
\end{figure*}

\begin{table}[t]
\centering
\scriptsize
\caption{Physical and intents parameters}\label{tab: exp}
\begin{tabular}{|ll|ll|}
\hline
\multicolumn{2}{|l|}{Physical Networks}          & \multicolumn{2}{l|}{Intents Parameters}                      \\ \hline
\multicolumn{1}{|l|}{edge node} & 200            & \multicolumn{1}{l|}{req num.}       & {[}1, 4{]}             \\ \hline
\multicolumn{1}{|l|}{edge link} & 758            & \multicolumn{1}{l|}{vir. node/link} & {[}2, 4{]}/{[}2, 4{]}  \\ \hline
\multicolumn{1}{|l|}{node CPU}  & {[}10,40{]}    & \multicolumn{1}{l|}{vir. CPU}       & {[}1, 2{]}             \\ \hline
\multicolumn{1}{|l|}{node RAM}  & {[}10,80{]}    & \multicolumn{1}{l|}{vir. RAM}       & {[}1, 4{]}             \\ \hline
\multicolumn{1}{|l|}{link bw}   & {[}400,1000{]} & \multicolumn{1}{l|}{bw/delay}       & {[}1,2{]}/{[}10,100{]} \\ \hline
\end{tabular}
\vspace{-0.5cm}
\end{table}

We extract the taxi GPS data within one hour with the number of taxis ranging from 1000 to 3000.
The location of each edge node or server is calculated based on the density of base stations of Shanghai telecom by the K-mean algorithm (K=200)~\cite{guo2020user}. Physical links within the VEC network are generated based on the Delaunay Triangulation algorithm~\cite{2020SciPy-NMeth}. As a result, there are 758 physical links with a 6.6852 km average distance between edge servers. The average distance between a user and the nearest edge station is 1.63 km.

The delays of wireless connection between users and base stations and wired network link are calculated as follows: 
\begin{equation}
    t = t_{wireless} + t_{wired}
\end{equation}
$t_{wireless} = W*\log_{2}{\frac{Sg_t}{N}}$
where the channel gain $g_t$ is $127 + 30 * \log(d)$, where $d$ is the distance between the user and local base station~\cite{niu2017}.
The channel bandwidth $W$ is set to 20 Mhz, the noise power $N$ is $2*10^{-13}$ Watt, and the wireless transmit power of vehicle S is 0.5 Watt.
The propagation time of wired links in milliseconds is calculated as $t_{wired} = 0.005 * d$, where $d$ in $km$ is the length of the direct optical cables.

An intent is generated and submitted at the same timestamp when the Taxi ID first appeared in the GPS data. The same intent is terminated when the Taxi ID is last shown in the data. Microservices are generated after the intent compiling (Fig.~\ref{intent-lifecycle}).
Fig.~\ref{fig: input} depicts the number of intent submissions and suspended events over time, and the total submissions and suspended event number of intent requests with various user numbers. A large number of intents were submitted by users before time 25. Table~\ref{tab: exp} illustrates the details of the physical network and intent parameters. Experimental results are reported based on the average of 10 randomly generated values.

\begin{figure*}[t]
\centering
\begin{subfigure}[t]{.23\linewidth}
	\centering
	\includegraphics[width=\linewidth, trim={0.2cm 0.3cm 0.2cm 0.2cm}, clip]{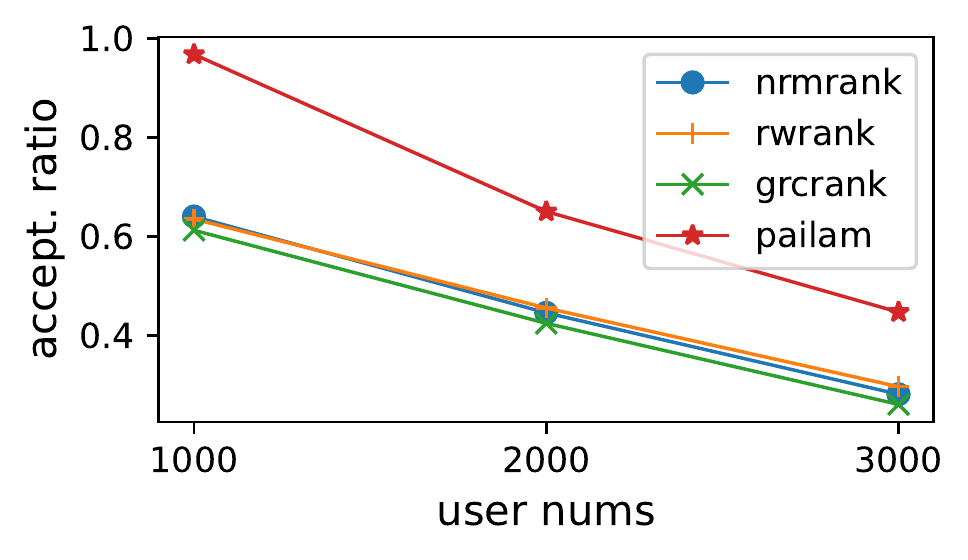}
	\caption{acceptance ratio with various user amounts}
	\label{fig: noderank-totalratio}
\end{subfigure}
\hspace{0.1em}
\begin{subfigure}[t]{.23\linewidth}
	\centering
	\includegraphics[width=\linewidth, trim={0.2cm 0.3cm 0.2cm 0.2cm}, clip]{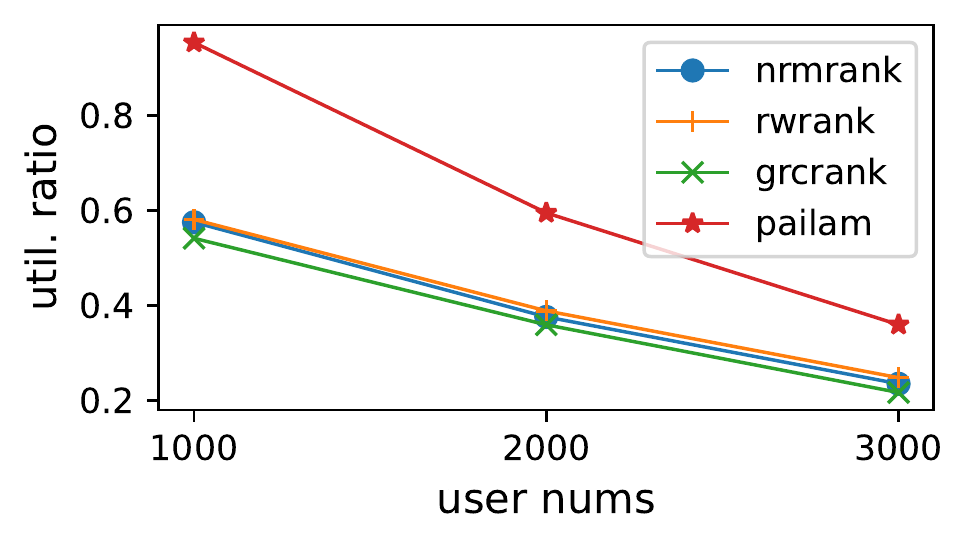}
	\caption{utilization ratio with various user amounts}
	\label{fig: noderank-util}
\end{subfigure}
\hspace{0.1em}
\begin{subfigure}[t]{.23\linewidth}
	\centering
	\includegraphics[width=\linewidth, trim={0.2cm 0.3cm 0.2cm 0.2cm}, clip]{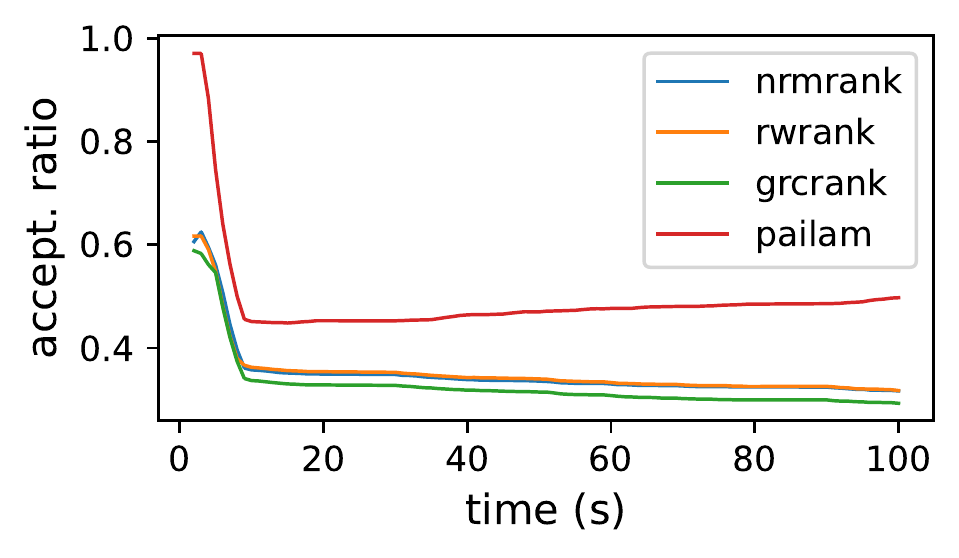}
	\caption{acceptance ratio with 3000 users along time}
	\label{fig: noderank-ratio}
\end{subfigure}
\hspace{0.1em}
\begin{subfigure}[t]{.23\linewidth}
	\centering
	\includegraphics[width=\linewidth, trim={0.2cm 0.3cm 0.2cm 0.2cm}, clip]{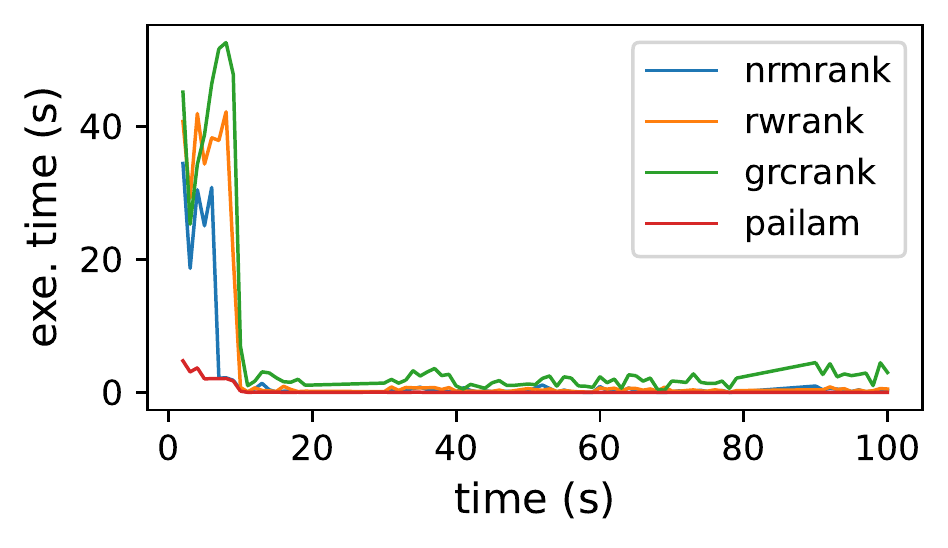}
	\caption{execution time with 3000 users along time}
	\label{fig: noderank-exe}
\end{subfigure}
\hspace{0.1em}
\caption{Installation performance comparisons with different online algorithms}
\label{fig: exp-noderank}
\vspace{-0.3cm}
\end{figure*}

\begin{figure*}[th]
\centering
\begin{subfigure}[t]{.23\linewidth}
	\centering
	\includegraphics[width=\linewidth, trim={0.2cm 0.3cm 0.2cm 0.2cm}, clip]{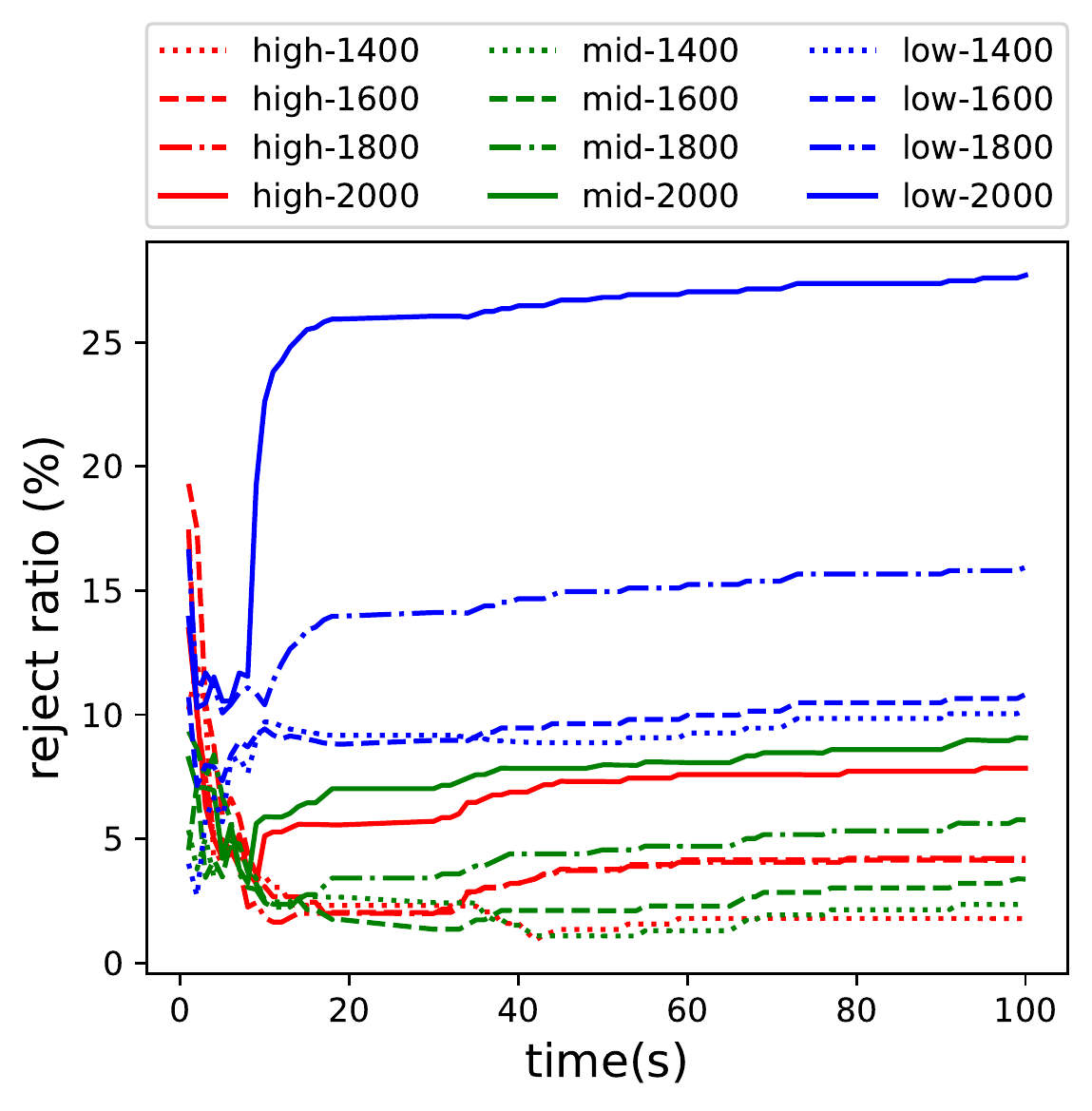}
	\caption{accept. ratio with various priorities}
	\label{fig: exp-pri}
\end{subfigure}
\hspace{0.1em}
\begin{subfigure}[t]{.23\linewidth}
	\centering
	\includegraphics[width=\linewidth, trim={0.2cm 0.3cm 0.2cm 0.2cm}, clip]{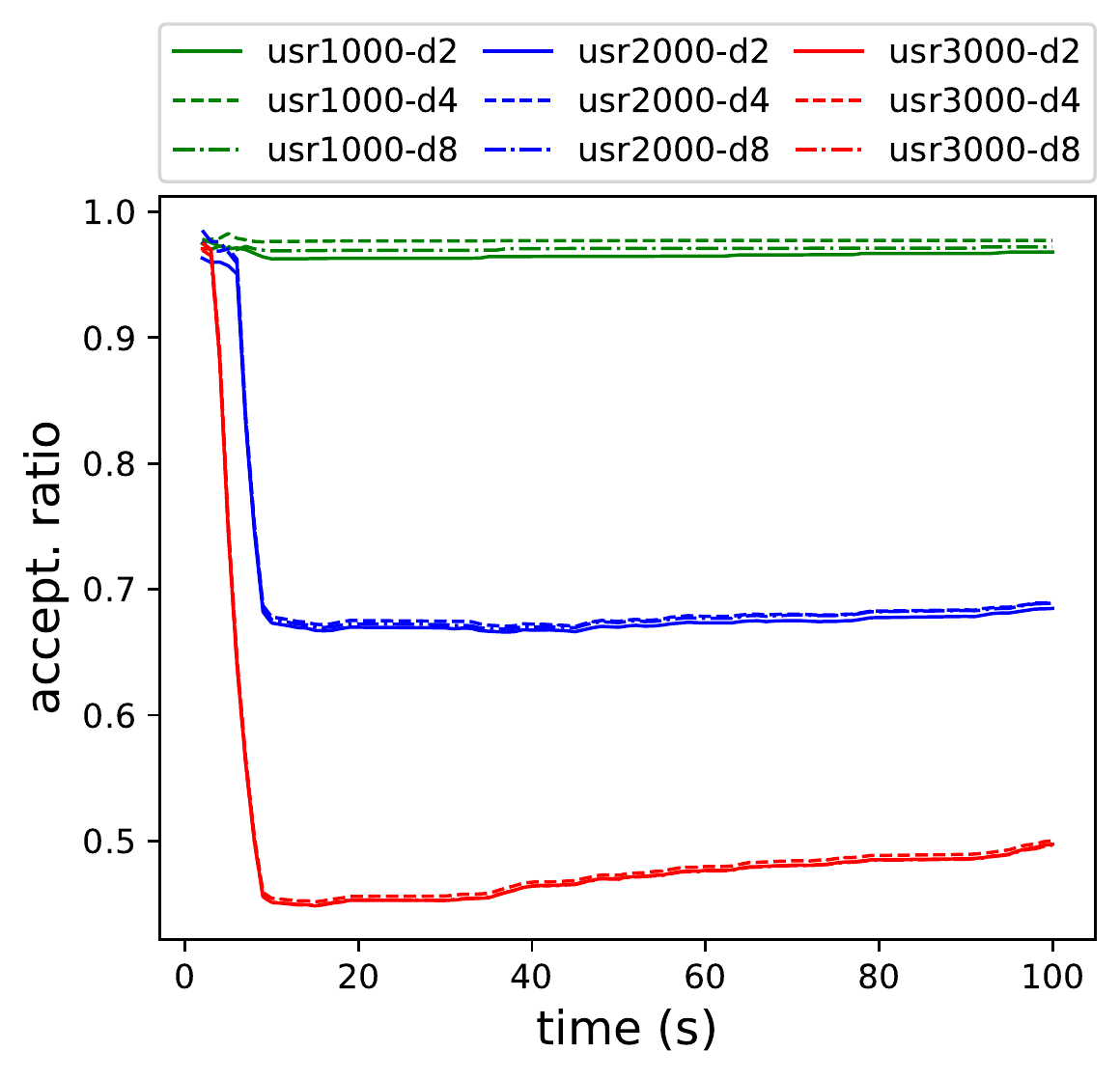}
	\caption{accept. ratio with various search depths} 
	\label{fig: exp-depth}
\end{subfigure}
\hspace{0.1em}
\begin{subfigure}[t]{.23\linewidth}
	\centering
	\includegraphics[width=\linewidth, trim={0.2cm 0.3cm 0.2cm 0.2cm}, clip]{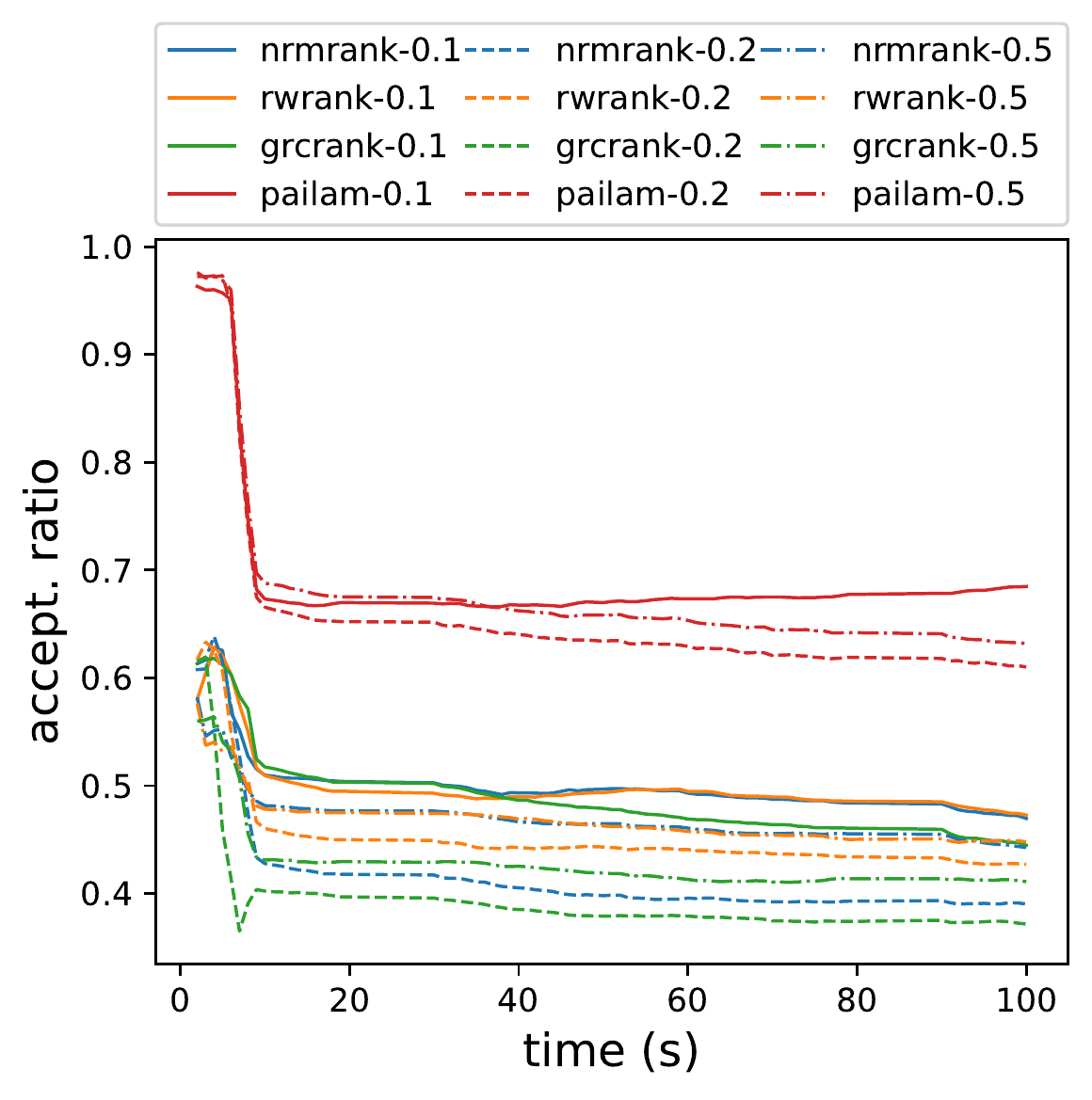}
	\caption{user-location constraints with $2000$ users} 
	\label{fig: exp-loc}
\end{subfigure}
\hspace{0.1em}
\caption{Installation performance with various amount of users for different priorities, search depth, and user location constraints}
\label{fig: exp-results1}

\vspace{-0.4cm}
\end{figure*}

\subsection{Results Analysis}
We evaluate the proposed PAI and LAM algorithms (\textit{pailam}) described in Sections~\ref{sec: PIA} and \ref{sec: LAM} in various aspects, namely intent acceptance ratio, resource utilization, and execution time. 
Figure~\ref{fig: exp-noderank} illustrates the performance comparisons in terms of acceptance ratio, utilization ratio, and execution time with various numbers of users. The ratio of location constraint requests to all requests is 0.1. It shows that our proposed LAM algorithm can efficiently install delay-sensitive requests with and without location constraints. LAM significantly increases the intent acceptance ratio (Fig.~\ref{fig: noderank-totalratio}) and utilization (Fig.~\ref{fig: noderank-util}) by up to $58$\%-$71$\% and $66$\%-$76$\%, respectively, compared with other online installation algorithms.
By considering the entire physical network for mapping, the execution time of baseline VNE algorithms is too high to suit the large-scale intent installation (Fig.~\ref{fig: noderank-exe}).
Due to the appropriate candidate selections, the processing time of LAM is decreased by up to $95$\% compared with other online algorithms.
We further evaluate our proposed intent framework and \textit{pailam} algorithm with various parameters, including the intent priority, search depth for the node mapping candidates, and the ratio of requests with location constraints (Fig.~\ref{fig: exp-results1}).

\textbf{Intent priority}: Figure~\ref{fig: exp-pri} shows the reject ratio along the time, in which high, mid, and low-level intent priorities are evenly generated. Compared to the mid-priority intents, the reject ratio of high-priority intents are significantly smaller. Both mid and high-priority intents can be maintained at a high acceptance ratio ($0.923$-$0.979$ and $0.931$-$0.979$). Intents with low priority ($0.741$-$0.907$) are rejected when one of the requests is not satisfied for higher-priority intents.

\textbf{Allocation search depth}: 
With a fixed location constraints ratio of 0.1, we examine the acceptance ratio of \textit{pailam} with various search depths for node mapping (Fig. \ref{fig: exp-depth}). The difference in acceptance ratio between depths $2$, $4$, and $8$ is insignificant among the different numbers of users. With a larger group of candidates, the depth $d$$=$$4$ has the best performance in all scenarios. However, the increase in algorithm execution time is considerably greater than the increase in performance compared to $d$$=$$2$. When $d$$=$$8$, the acceptance ratio decreases slightly because the larger group of candidates may lead to more mapping rejections.

\textbf{Location constraints}: 
Figure \ref{fig: exp-loc} illustrates the acceptance ratio when the percentage of user location-related requests varies for 2000 users. \textit{pailam} uniformly and significantly outperforms other online algorithms in all scenarios. The acceptance ratio is the highest in the scenario where $10$\% of total intents are location-related (pailam-0.1). The acceptance ratio of location-related requests may be higher than the non-location-related requests when the computing and networking resources of $n(v)$ locations with mapping constraints are sufficiently large. For example, when the user's onboard processing resources and its network connections to edge servers are sufficiently large (0.5), the acceptance ratio is higher than the $0.2$ scenario.

\textbf{Link remapping and intent reinstallation}: 
Suspended events are triggered along the time due to user mobility (Fig.~\ref{fig: input-suspend}). Compared to baseline VNE algorithms, the proposed algorithm does not directly map the suspended requests as if they are new submissions. As a result, it largely reduces the execution time and the cost of VM or container migrations for intent reinstallation~\cite{he2022camig}. This is because the operating cost of path rerouting is significantly smaller than live migration~\cite{he2019mig}.

\section{Intent-Based Computing Prototype}
In this section, we showcase an Intent-Based Vehicular Edge Computing prototype based on the SDN controller (ONOS) and Mininet Emulation platform.\footnote{You can find a demo at \url{https://youtu.be/ZXBXdxug_x4}} Virtual node (VM and container) mapping and virtual link embedding are controlled by the intent-based framework. We generated a series of events to validate the feasibility, availability, and flexibility of our proposed system (Fig. \ref{fig: prototype}).
With a 100 priority (low-level), $Intent1$ is compiled into one request with $v^1_1$, $v^1_2$ and $v^1_3$ and links $v^1_1$-$v^1_2$ and $v^1_1$-$v^1_3$. The requirement of each node is 2 vCPUs and 4 GB RAM and the location constraint of $v^1_1$ is $user1$. Each virtual link requires 20 Mbps of bandwidth and 30 ms of latency. With a 200 priority (mid-level), $Intent2$ is compiled into two requests with virtual link $v^2_1$-$v^2_2$ and $v^2_1$-$v^2_3$, respectively. The requirement of each node is 2 vCPUs and 4 GB RAM and the location constraint of $v^2_2$ is $EDC1$. The link requirement is 20 Mbps and 100 ms.

\begin{figure*}[t]
    \centering
    \includegraphics[width=1\linewidth]{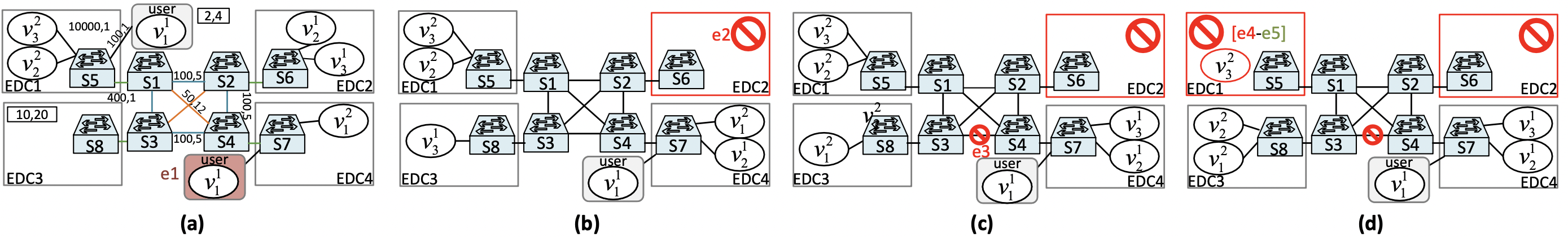}
    \caption{Emulation scenarios. The numbers over the links show the bandwidth and delay, i.e. 400,1.}
    \label{fig: prototype}
    \vspace{-0.3cm}
\end{figure*}

\begin{figure*}[t]
\centering
\begin{subfigure}{.23\linewidth}
	\centering
	\includegraphics[width=\linewidth, trim={0.2cm 0.3cm 0.2cm 0.2cm}, clip]{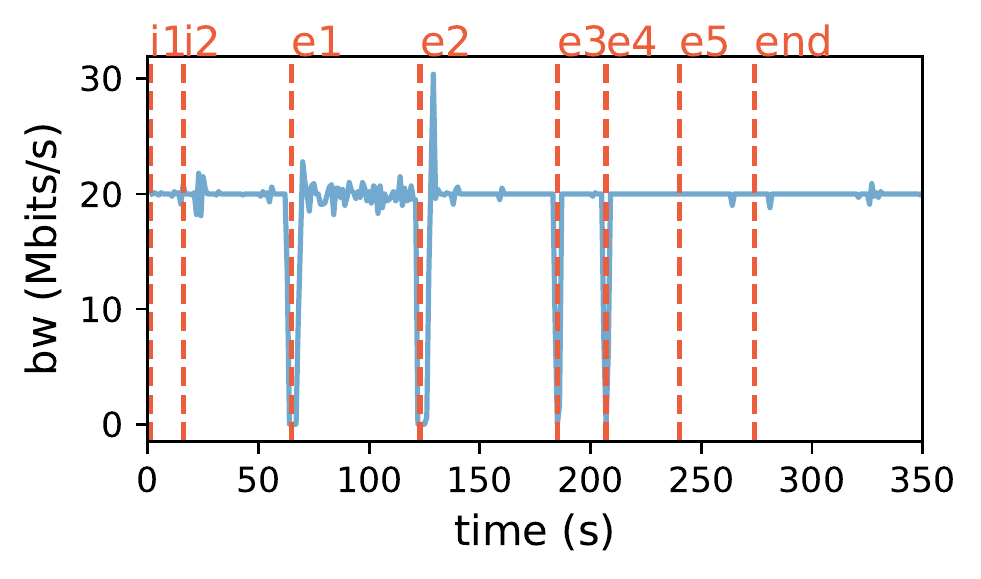}
	\caption{$v^1_1$ and $v^1_2$}
	\label{fig: project-bw1}
\end{subfigure}
\hspace{0.1em}
\begin{subfigure}{.23\linewidth}
	\centering
	\includegraphics[width=\linewidth, trim={0.2cm 0.3cm 0.2cm 0.2cm}, clip]{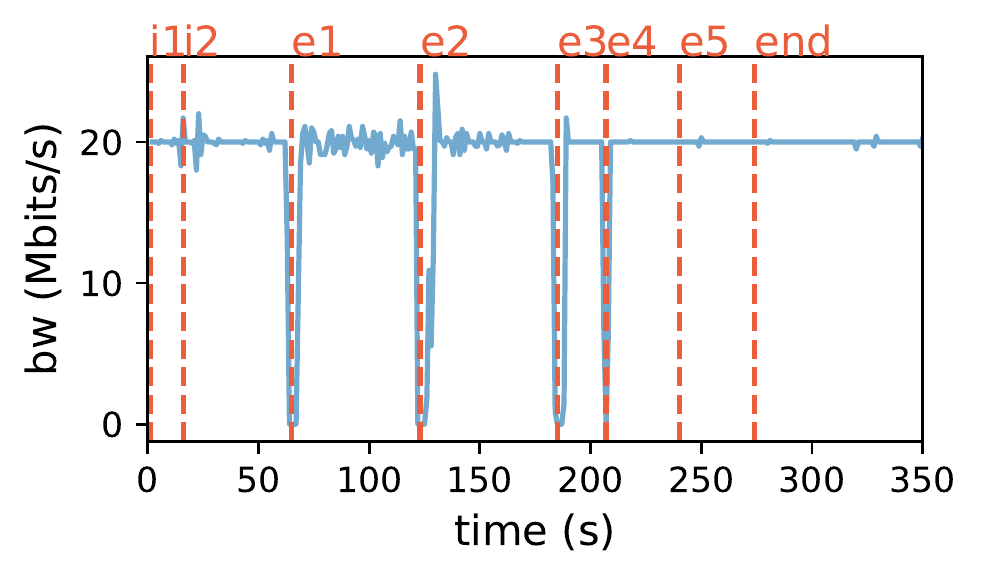}
	\caption{$v^1_1$ and $v^1_3$}
	\label{fig: project-bw2}
\end{subfigure}
\hspace{0.1em}
\begin{subfigure}{.23\linewidth}
	\centering
	\includegraphics[width=\linewidth, trim={0.2cm 0.3cm 0.2cm 0.2cm}, clip]{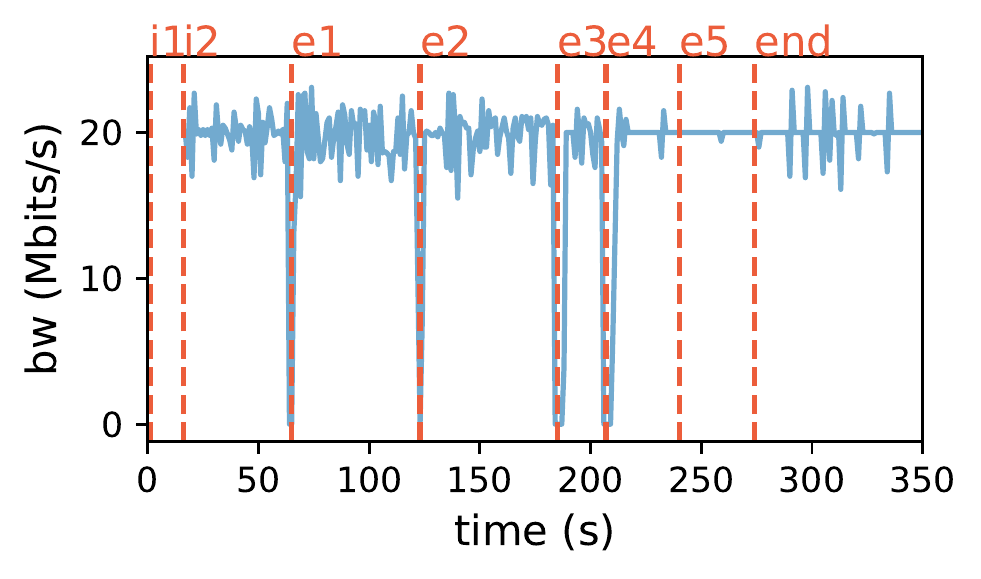}
	\caption{$v^2_1$ and $v^2_2$}
	\label{fig: project-bw3}
\end{subfigure}
\hspace{0.1em}
\begin{subfigure}{.23\linewidth}
	\centering
	\includegraphics[width=\linewidth, trim={0.2cm 0.3cm 0.2cm 0.2cm}, clip]{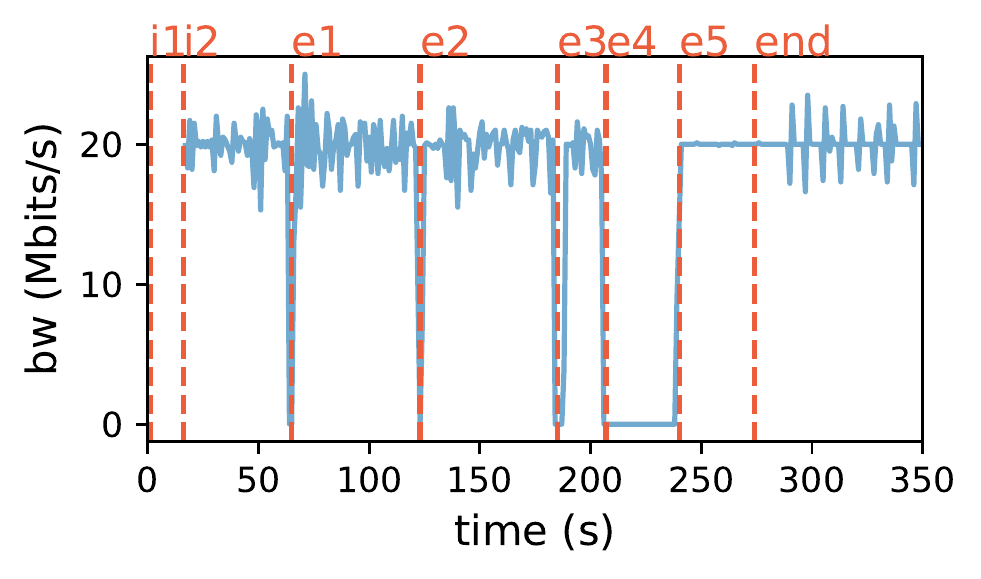}
	\caption{$v^2_1$ and $v^2_3$}
	\label{fig: project-bw4}
\end{subfigure}
\hspace{0.1em}
\caption{Emulation performance in bandwidth along the time}
\label{fig: project-bw}
\vspace{-0.3cm}
\end{figure*}

\begin{figure*}[t]
\centering
\begin{subfigure}{.23\linewidth}
	\centering
	\includegraphics[width=\linewidth, trim={0.2cm 0.3cm 0.2cm 0.2cm}, clip]{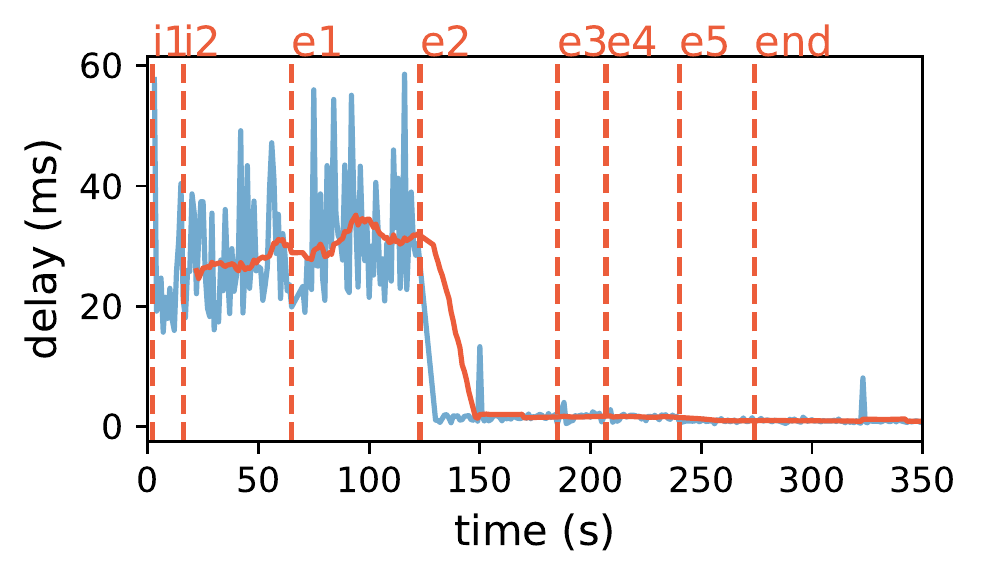}
	\caption{$v^1_1$ and $v^1_2$}
	\label{fig: project-delay1}
\end{subfigure}
\hspace{0.1em}
\begin{subfigure}{.23\linewidth}
	\centering
	\includegraphics[width=\linewidth, trim={0.2cm 0.3cm 0.2cm 0.2cm}, clip]{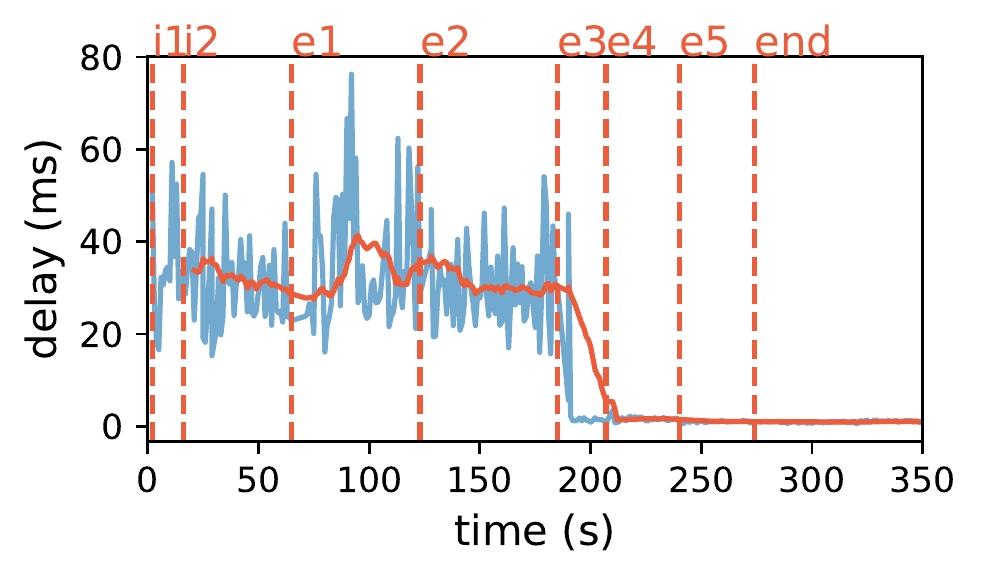}
	\caption{$v^1_1$ and $v^1_3$}
	\label{fig: project-delay2}
\end{subfigure}
\hspace{0.1em}
\begin{subfigure}{.23\linewidth}
	\centering
	\includegraphics[width=\linewidth, trim={0.2cm 0.3cm 0.2cm 0.2cm}, clip]{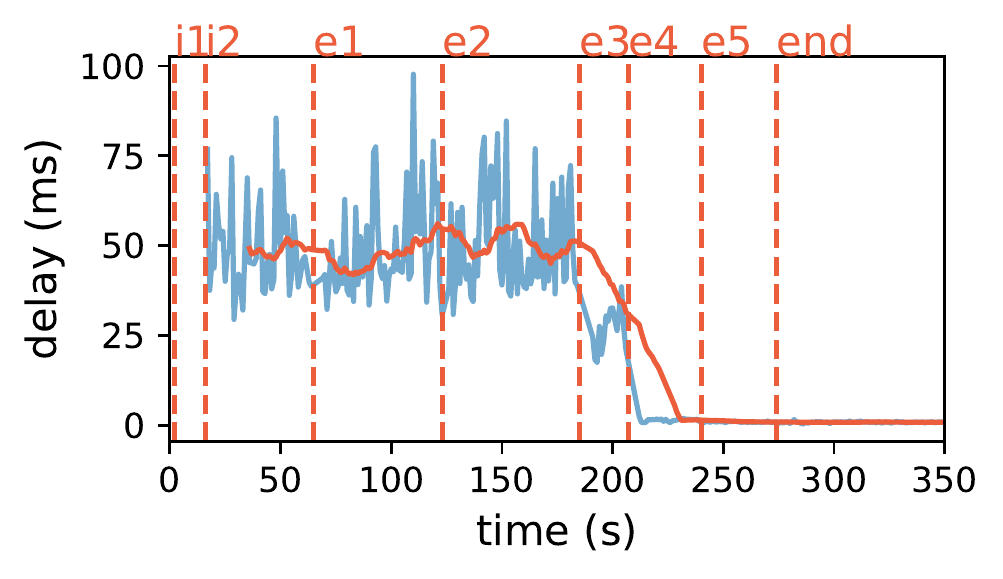}
	\caption{$v^2_1$ and $v^2_2$}
	\label{fig: project-delay3}
\end{subfigure}
\hspace{0.1em}
\begin{subfigure}{.23\linewidth}
	\centering
	\includegraphics[width=\linewidth, trim={0.2cm 0.3cm 0.2cm 0.2cm}, clip]{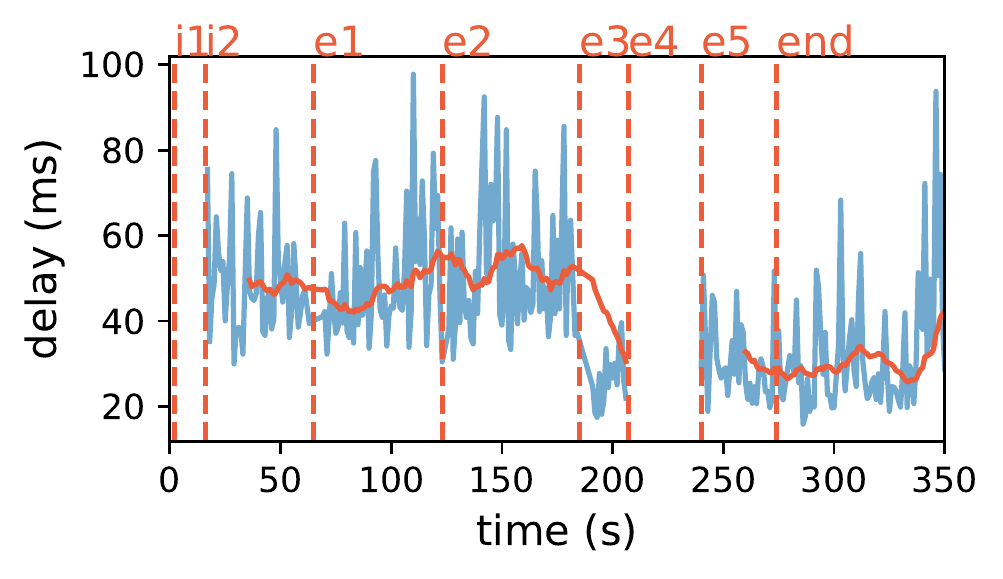}
	\caption{$v^2_1$ and $v^2_3$}
	\label{fig: project-delay4}
\end{subfigure}
\hspace{0.1em}
\caption{Emulation performance in delay and its rolling average along the time}
\label{fig: project-delay}
\vspace{-0.4cm}
\end{figure*}

At time \textit{i1} and \textit{i2}, \textit{intent1} and \textit{intent2} are submitted and installed, respectively. At \textit{e1}, $user1$ who is connected to EDC1 moves to a new position and gets connected to EDC4. As a result, a mobility event is raised to check the installation of \textit{intent1} (Fig. \ref{fig: prototype}(a)). At \textit{e2}, EDC2 goes down. $v^1_2$ and $v^1_3$ are reallocated to satisfy \textit{intent1} (Fig. \ref{fig: prototype}(b)). At time \textit{e3}, the link between S3 and S4 goes down (Fig. \ref{fig: prototype}(c)). Between \textit{e4} and \textit{e5}, request $v^2_1$-$v^2_3$ of \textit{intent2} failed due to EDC1 is down and $v^2_3$ location constraint cannot be satisfied (Fig. \ref{fig: prototype}(d)). At $e5$, EDC1 is up again, and request $v^2_1$-$v^2_3$ is reinstalled. As shown in Fig.~\ref{fig: project-bw} and Fig.~\ref{fig: project-delay}, the prototype application can react to the computing, networking, and mobility events, satisfy the intents requirements and manage the life-cycle of intents efficiently.

\section{Related work}
Virtual network embedding (VNE) \cite{Fischer2013} refers to the embedding of a virtual network in a substrate network. 
To provide custom user-defined end-to-end guaranteed services to end users, there are different problems that are addressed in this problem domain, such as optimal resource allocation, self-configuration, and organization of the network. Yu \textit{et al.}~\cite{Yu2008} showed an approach that combines path splitting, path migration, and customized embedding algorithms to enable a substrate  network to satisfy a larger mix of virtual networks. Dietrich~\textit{et al.}~\cite{Dietrich2013} addressed the problem of multi-provider VNE with limited information disclosure. 
EPVNE~\cite{Li2019} is a heuristic algorithm that reduces the cost of embedding the Virtual Network (VN) request and increases the VN request acceptance ratio. Hejja and Hesselbach~\cite{Hejja2018} proposed an online power-aware algorithm to solve the VNE problem using fewer resources and less power consumption with end-to-end delay as a constraint. Jinke~\textit{et~al.}~\cite{Jinke2013} proposed a VNE model, where high-priority users can get extra resources compared to low-priority users. Ogino~\textit{et al.}~\cite{Ogino2017} proposed a VNE method to minimize the total substrate resources required during substrate resource sharing among multiple priority classes. Nguyen~\textit{et al.}~\cite{Nguyen2020} proposed a node-ranking approach, and a parallel GA-based algorithm for the link mapping stage to solve the online VNE problem. DeepViNE~\cite{Dolati2019} is a Reinforcement Learning (RL)-based VNE solution to automate the feature selection. MUVINE~\cite{Thakkar2020} is an RL-based prediction model for multi-stage VNE among cloud data centers.

Recently, there has been some research to improve network functionalities with the help of SDN intents in different application scenarios. ONOS Intent Framework~\cite{campenella2019} indicates the IBN operations used in the ONOS SDN controller. Han~\textit{et al.}~\cite{Han2017} proposed an IBN management platform based on SDN virtualization  to automate the management and configuration of virtual networks. 
Cerroni~\textit{et al}.~\cite{Cerroni2017} proposed a reference architecture and an intent-based North Bound Interface (NBI) for end-to-end service orchestration across multiple technological domains, with a primary use-case being the infrastructure deployment on the Internet of Things (IoT). DISMI~\cite{Skoldstrom2017} is also proposed as an intent-based north-bound interface of a network controller. Addad~\textit{et~al.}~\cite{Addad2018} benchmarked the ONOS intent NBI  using  a  methodology  that  takes  into  consideration the interface access method, type of intent, and the number of installed intents. 
OSDF~\cite{Comer2018} is an SDN-based network programming framework that provides high-level APIs to be used by managers and network administrators to express network requirements for applications and policies for multiple domains. 
Sanvito~\textit{et~al.}~\cite{Sanvito2018} extended the ONOS Intent Framework enabling compiling multiple intents and re-optimizing their paths according to the network state based on flow statistics. Szyrkowiec~\textit{et~al.}~\cite{Szyrkowiec2018} proposed architecture for automatic intent-based provisioning of secure services in multi-layer IP, Ethernet, and optical networks while choosing the appropriate encryption layer.  Rafiq~\textit{et~al.}~\cite{Rafiq2020} enables IBN to make effective network resource utilization and minimize the maximum link capacity utilization.

In summary, existing VNE algorithms do not support allocating multiple virtual nodes to the same compute node, and VNRs are treated as individual requests. However, our work supports VEC applications compiled into multiple VNRs. Our approach also extends intents with location constraints and incorporates computing requirements, while satisfying users' QoS requirements. It can also handle the mobility aspect of the users/applications. Unlike existing VNE algorithms, our approach is suitable for intent installation with priorities.

\section{Conclusions and future work }
We proposed a novel intent framework to jointly orchestrate networking and computing requirements of applications based on user requirements in vehicular edge computing environments. Our proposed solution constantly monitors user requirements and dynamically reconfigures the system to satisfy the desired states of applications. It optimizes resource utilization and acceptance ratio of computing and networking requests with various priorities. Results show that our proposed framework outperforms state-of-the-art algorithms in terms of acceptance ratio, resource utilization, and execution time. We also provided a small-scale prototype to validate our proposed framework. In future work, we plan to fully implement our framework by extending the intent framework of the ONOS SDN controller and consider inherited periodic mobility patterns for intent re-installation and management.

\bibliographystyle{IEEEtran}
\bibliography{ref}
\end{document}